\documentclass[11pt]{article}

\usepackage[utf8]{inputenc}
\usepackage{jheppub,amsmath,amssymb,mathabx}
\usepackage{bm,bbm}
\usepackage{smartdiagram}
\usepackage{metalogo}
\usepackage{dtklogos}
\usepackage{forest}
\usepackage{lscape}
\usepackage{xcolor}
\usepackage{float}
\usetikzlibrary{arrows.meta, shapes.geometric, calc, shadows}

\usepackage{wrapfig}
\usepackage{lscape}
\usepackage{epstopdf}
\usepackage{bbding}
\usepackage{rotating}
\newcommand{\tn}[1]{\begin{turn}{90}#1\end{turn}}

\colorlet{mygreen}{green!75!black}
\colorlet{col1in}{purple!10}
\colorlet{col1out}{purple}
\colorlet{col2in}{mygreen!40}
\colorlet{col2out}{mygreen!50}
\colorlet{col3in}{blue!30}
\colorlet{col3out}{blue!40}
\colorlet{col4in}{mygreen!20}
\colorlet{col4out}{mygreen!30}
\colorlet{col5in}{blue!10}
\colorlet{col5out}{blue!20}
\colorlet{col6in}{blue!20}
\colorlet{col6out}{blue!30}
\colorlet{col7out}{orange}
\colorlet{col7in}{orange!50}
\colorlet{col8out}{orange!40}
\colorlet{col8in}{orange!20}
\colorlet{linecol}{blue!60}
\definecolor{col9in}{HTML}{E8D4FE}
\definecolor{col9out}{HTML}{CDBAFE}
\definecolor{colinGal}{rgb}{0.69,0.16,0.53}
\definecolor{coloutGal}{rgb}{0.69,0,0.53}
\definecolor{MGout}{HTML}{5614FC}
\definecolor{MGin}{HTML}{68A9F8}
\definecolor{NLout}{HTML}{6DD6F7}
\definecolor{NLin}{HTML}{BBF3FB}

\allowdisplaybreaks[3]

\usepackage{amsmath, amsfonts, amsthm, amssymb, graphicx, xcolor}
\usepackage{hyperref}
\hypersetup{
    colorlinks,
    linkcolor={bg2},
    citecolor={bg2},
    urlcolor={bg2}
}
\usepackage{subfigure}

\definecolor{color1}{rgb}{0.80,0.00,0.60}
\definecolor{bg2}{rgb}{0.00,0.40,0.99}

\newcommand{\mpl}{M_{\mathrm{Pl}}}

\newcommand{\U}{\mathcal{U}}
\newcommand{\K}{\mathcal{K}}
\newcommand{\ie}{{\it i.e.,}\ }
\newcommand{\eg}{{\it e.g.,}\ }

\renewcommand{\mpl}{M_{\rm Pl}}

\renewcommand{\[}{\left[}

\renewcommand{\L}{\mathcal L}

\def\nn{\nonumber}

\def\d{\mathrm{d}}

\def\ba{\begin{eqnarray}}
\def\ea{\end{eqnarray}}

\def\L{\mathcal{L}}
\def\K{\mathcal{K}}

\def\E{\mathcal{E}}

\def\stu{St\"uckelberg }

\def\d{\mathrm{d}}
\def\mn{_{\mu \nu}}
\def\ab{_{\alpha \beta}}

\def\mupn{^\mu_{\ \nu}}

\def\({\left(}
\def\){\right)}

\def\mpl{M_{\rm Pl}}
\def\p{\partial}
\def\ien{{\em i.e.}}

\def\L{\mathcal{L}}
\def\K{\mathcal{K}}

\def\nn{\nonumber}

\def\d{\mathrm{d}}

\def\mn{_{\mu \nu}}
\def\mupn{^\mu_{\, \nu}}

\def\({\left(}
\def\){\right)}

\def\nn{\nonumber}
\def\p{\partial}

\usepackage[normalem]{ulem}
\usepackage{chemarrow}

\title{Massive Gravity @ 15}
\author[a,b,c]{Claudia de Rham}
\affiliation[a]{Abdus Salam Centre for Theoretical Physics, Imperial College, London, SW7 2AZ, UK}
\affiliation[b]{CERCA, Department of Physics, Case Western Reserve University, 10900 Euclid Ave, Cleveland, OH 44106, USA}
\affiliation[c]{Perimeter Institute for Theoretical Physics, 31 Caroline St N, Waterloo, ON, N2L 2Y5, Canada}
\emailAdd{c.de-rham@imperial.ac.uk}

\preprint{{\footnotesize Imperial--TP--2026--cdr--3~~~~~~~~ }}
\date{}

\abstract{
Massive gravity\footnote{Invited contribution to FQxI's upcoming collection of reviews about the biggest open problems in foundational physics.} 
is one of the most natural proposals for modifying gravity at large distance scales. First proposed to tackle the cosmological constant problem, it has the potential to simultaneously maintain theoretical consistency and agreement with current observational constraints. The development of Lorentz-invariant ghost-free massive gravity resolves long-standing obstacles associated with interacting massive spin--2 fields. Central to this development is a highly constrained nonlinear interaction structure that propagates exactly five degrees of freedom. The same interaction structure raises the strong-coupling scale to its maximal value, which organizes the theory as an effective field theory. Phenomenological consistency with local tests of gravity is ensured by nonlinear screening through the Vainshtein mechanism, which is automatically built in. A precise decoupling limit may be defined which singles out the dominant interactions at the $\Lambda_3$ scale, leading to generalised classes of Galileon theories.  Recent developments have shown how to formulate the theory in a manifestly well-posed way both within and beyond this decoupling limit. From an effective field theory perspective, unitarity, analyticity, and causality impose powerful constraints that strongly single out the ghost-free $\Lambda_3$ theory as a distinguished infrared realisation of massive gravity. We further discuss these results in light of recent consistency analyses, which we put in context. 
Interestingly, massive gravity has recently been invoked in discussions of black-hole entanglement entropy and spacetime regions known as `islands'.
The theory is also shown to emerge as an exactly solvable $T \bar T$ deformation in both two dimensions and for special cases in higher dimensions, opening up valuable insights into its UV behaviour. 
Massive Gravity thus serves as a valuable theoretical laboratory for exploring the interplay between phenomenology and UV completions in infrared modifications of General Relativity.  
}

\begin{document}

\maketitle

\section{Introduction} 

General Relativity (GR) is one of the great pillars of twentieth-century physics. Unlike the other great pillar, quantum mechanics, its experimental verification was somewhat slow, in large part because its most novel signatures are at astrophysical and cosmological scales. Furthermore, its most remarkable predictions, black holes and the Big-Bang singularity, were long thought to be no more than mathematical curiosities having no direct physical relevance. Today it stands as one of the most stringently tested theories in fundamental physics, with black holes very much at the forefront, and is continuously being confronted with experimental and observational scrutiny across an extraordinary range of physical scales. 
In the weak-field regime, laboratory experiments and solar-system tests have confirmed GR predictions through measurements of light deflection, time-delay experiments, perihelion precession, and tests of the equivalence principle, all within the precision of current experiments \cite{Will:2014kxa,Will:2018bme}. Since these are weak-field tests, they demonstrate consistency both with Newtonian gravity in the non-relativistic limit and also with the spin--2 nature of gravity in the weak-field but relativistic limit. In particular, the famous prediction of the bending of light around the sun is most naturally reproduced by recognising the spin--2 nature of gravity.

At larger distances, astrophysical and cosmological observations probe the more non-trivial dynamical and relativistic regimes where Newtonian gravity is far from valid. These include binary pulsars \cite{Hulse:1974eb}, which also provide indirect tests of the spin--2 nature of gravity, gravitational lensing, and, more recently, the direct detection of gravitational waves from compact binary mergers of black holes and neutron stars by the LIGO--Virgo-KAGRA Collaboration \cite{Abbott:2016blz}. The latter provides an indirect test of GR in the strong-curvature regime through the gravitational waves emitted by coalescing black holes, a domain in which the weak-field spin-2 description no longer applies. 
The recent imaging of supermassive black holes by the Event Horizon Telescope \cite{EventHorizonTelescope:2019dse,EventHorizonTelescope:2022wkp} has given us an observational window on the spacetime geometry near the event horizon, further constraining our understanding of strong-field gravity.

Despite these remarkable empirical successes, we cannot be fully confident that General Relativity is the end of the story. The first real doubt comes from cosmology, a field which in the last $\approx 30$ years has transformed from a hopeful but undeveloped and exploratory field, to what is now a precise science, capable of successfully fitting large swathes of data, and having the luxury of a `Standard Model'.
The main challenge for the otherwise successful cosmic concordance model is the late-time accelerated expansion of the Universe. This phenomenon is inferred from observations of Type~Ia supernovae \cite{SupernovaSearchTeam:1998fmf,Perlmutter:1999jt,Planck:2018vyg}, in combination with data from the cosmic microwave background, baryon acoustic oscillations, and large-scale structure surveys \cite{Planck:2018vyg}.

This acceleration is most economically described in GR by the addition of a cosmological constant to Einstein's equations. The constant is best understood as a contribution to the vacuum energy density, which naturally induces a stress-energy tensor proportional to the metric. A less economical approach is to promote this stress-energy to a dynamical form of energy, known simply as dark energy. This is, at best, a proxy for our ignorance. It is a useful tool for describing the dynamics of that component, but it does not necessarily correspond directly to any well-defined fundamental theory.

Despite its mathematical simplicity, the introduction of a cosmological constant, while consistent with current observations, raises problems of its own. The observed value of the cosmological constant needed is many orders of magnitude smaller than the naïve expectations from quantum field theory. This is true essentially from any source of vacuum energy we might imagine (except perhaps neutrinos). This tension, commonly referred to as the cosmological constant problem, has long been recognised as one of the deepest unresolved problems in theoretical physics \cite{Weinberg:1988cp} and is in part so severe that many practitioners have chosen to gloss over or even declare the prediction of its value from fundamental principles as impossible.

These difficulties strongly motivate the exploration of alternatives to the standard cosmological paradigm. These alternatives could come in the form of more intricate and dynamical descriptions of dark energy, such as quintessence models, or more major modifications which can be summarised under the umbrella of ``infrared modifications of gravity". Indeed, just as GR is understood to be the leading term in an infinite series of ultraviolet (UV) corrections (which become relevant at or below the Planck scale), one may also contemplate the possibility that it requires  infrared  (IR) modifications. 
Here, IR indicates that the deviations from GR become significant only at large (spacetime) distances or potentially also at low curvatures. All such alternatives will, of course, be subject to the stringent requirements of consistency with local tests of gravity. Within this broad landscape, massive gravity occupies a rather unique and theoretically special place. In massive gravity theories, the graviton itself is endowed with a small but nonzero mass, whose origin is some potential symmetry breaking mechanism, which may in practice have little to do with the Higgs mechanism. The very existence of a mass for the graviton leads to a modification of gravitational interactions at scales associated with the graviton's Compton wavelength. This length scale may be easily chosen to be of the order of current cosmological distances without contradicting any known test of gravity. This opens the possibility that vacuum energy itself may gravitate differently, with the influence of long wavelength fluctuations filtered out, a mechanism often referred to as degravitation \cite{Arkani-Hamed:2002ukf,Dvali:2007kt}.

Giving a mass to the graviton is by no means an innocent thing.
Such a deformation necessarily alters the number of propagating degrees of freedom relative to GR (at least in models which preserve Lorentz invariance), and hence produces more than a slight modification of the overall dynamics. This, in turn, is combined with theoretical challenges, including the need to ensure that additional polarisations are stable.

If massive gravity theories are to be phenomenologically viable, 
then there must be an explanation of why we have not seen the influence of these additional polarisations already. This is provided by the Vainshtein mechanism, which is an example of a nonlinear screening mechanism \cite{Vainshtein:1972sx}. The Vainshtein mechanism dynamically suppresses deviations from GR in regions of high density or curvature \cite{Vainshtein:1972sx,Babichev:2013usa}. Vainshtein constructed his original arguments at a time when a ghost-free massive gravity theory was not known\footnote{In the context of this review, a “ghost” refers to a dynamical degree of freedom whose kinetic term has the wrong sign, thereby causing arbitrarily rapid instabilities while remaining within the regime of validity of the theory.}. Now, more than 15 years ago, a ghost-free realisation of Lorentz-invariant massive gravity was discovered that eliminates the Boulware--Deser ghost \cite{Boulware:1972zf} by a specific choice of nonlinear interactions \cite{deRham:2010ik,deRham:2010kj,Hassan:2011hr}. The way in which the Vainshtein mechanism is realised in these theories is best understood through a particular decoupling limit \cite{Babichev:2013usa}.
Subsequent developments have shown how massive gravity can be written in a number of different formalisms, in each of which the ghost issue has been clarified. Significant focus has been on cosmological solutions, gravitational-wave phenomenology, and the theoretical constraints imposed by the consistency, causality, and analyticity of effective field theory \cite{deRham:2014zqa}.

In what follows, we provide an overview of the current status of Lorentz-invariant massive gravity, with some discussion of recent developments and open problems. Our exposition largely mirrors the logical progression of the subject as laid out in the review article \cite{deRham:2014zqa} and the book \cite{deRham:2023byw}, where the reader can find a much more extensive discussion and further references.

\section{The (Old) Cosmological Constant Problem}

Over the past several decades, cosmology has developed into a precise science, with a wide array of independent cosmological probes providing information on both the early and late Universe. 
Although not always mutually consistent, these data have, with increasing precision, largely converged towards a consistent `Standard Model of Cosmology', \ie $\Lambda$CDM. Measurements of the Cosmic Microwave Background (CMB) temperature and polarisation anisotropies provide the most detailed information on the early Universe. The CMB alone constrains the spatial geometry, primordial fluctuations, and the total energy budget \cite{Planck:2018vyg}. Type~Ia supernovae directly map the late-time expansion history \cite{SupernovaSearchTeam:1998fmf,Perlmutter:1999jt} and provide complementary observations which pin down the expansion history. Baryon acoustic oscillations and large-scale structure surveys trace the growth of the original primordial fluctuations across a wide range of redshifts to late times.
The recent direct detection of gravitational waves has offered an independent probe of cosmology through standard sirens, providing a novel handle on cosmic distances and the expansion rate \cite{Abbott:2017xzu}. Taken together, these diverse measurements form a coherent and largely consistent framework that provides compelling evidence for a late-time accelerated expansion of the Universe.

As already noted, within GR, by far the easiest explanation for this acceleration is a cosmological constant~$\Lambda$. At the level of the Einstein equations, $\Lambda$ corresponds to a uniform vacuum energy density $T^0{}_0$ that sources an asymptotically de~Sitter expansion at late times. From a purely phenomenological standpoint, this description is extraordinarily successful: The $\Lambda$CDM model (which includes a Cosmological Constant $\Lambda$ and Cold Dark Matter) provides an excellent fit to current observational data across all accessible scales and epochs, with no statistically compelling evidence for departures from a constant dark energy equation of state.

\subsection{Theoretical Challenge}

From the perspective of building a theoretical framework, the cosmological constant presents a profound challenge. In many ways its interpretation as a vacuum energy density arising from quantum fluctuations of the matter fields gives a satisfying explanation.
The problem is that the observed value of $\Lambda$ is extraordinarily small compared to any expected QFT calculation. Expressed in terms of an energy density, the required `dark energy' density is of the order of
\begin{equation}
\rho_\Lambda \sim 10^{-120} M_{\mathrm{Pl}}^4 ,
\end{equation}
where $M_{\mathrm{Pl}}$ denotes the reduced Planck mass. 
A simple one-loop calculation would imply a contribution to the vacuum energy of a particle of mass $m$ and spin $s$ of the form
\begin{equation}
\Delta \rho_\Lambda \sim (-1)^{2s} (2s+1) m_s^4 \, .
\end{equation}
Without any tuning, a natural explanation of the cosmological constant would require the masses of all particles to be less than $m_s \lesssim 10^{-30} \mpl \sim 10^{-3}{\rm eV}$. Every known massive particle, other than the neutrino, is considerably more massive than this. The smallness of $\rho_\Lambda$ therefore represents a deep tension between quantum field theory, gravity, and cosmology.\\

This tension is now referred to as the \emph{old cosmological constant problem}: the question of why vacuum energy contributions from quantum fields do not gravitate with their expected magnitude \cite{Weinberg:1988cp}. 
The most credible physical explanation is that a symmetry intervenes to resolve the issue and give a technically natural explanation of why this parameter is small.  
In particular, if the symmetry of the theory were to increase when one of its parameters vanished, then the radiative corrections to that parameter would be proportional  to the parameter itself and this parameter could then be small without suffering from large quantum corrections. In other words, some parameters can be  stable against quantum corrections when protected by an (approximate) symmetry, and setting such a parameter to be small is therefore technically natural \cite{tHooft:1979rat}.   
Despite extensive effort, no compelling and phenomenologically successful symmetry principle or dynamical mechanism has been identified within GR. The most solid theoretically is supersymmetry, which plays off the fact that fermions contribute with the opposite sign to bosons.

\subsection{Supersymmetry}

In pure supersymmetric theories, we can then have a sum rule of the form 
\begin{equation}
\sum_s (-1)^{2s} (2s+1) m_s^4 =0 \, ,
\end{equation}
causing the vacuum energy to vanish. The problem with this argument is phenomenological, clearly supersymmetry is broken in the real world, at scales above those of the Standard Model of Particle Physics. When the electron already gives a contribution that is expected to be a factor $10^{34}$ greater than the observed value, it is hard to see how the symmetry that comes into play on a higher energy scale helps. Another possible symmetry often considered is scalar (or conformal) invariance, which is similarly unsuccessful \cite{Weinberg:1988cp}. \\

A more recent take is summarised as the \emph{new cosmological constant problem}, which has to do with the apparent coincidence between the onset of cosmic acceleration and the present cosmological epoch. From this point of view, the puzzle is not only why $\Lambda$ is small, but also why its effects become cosmologically relevant precisely at late times. Of course, this could just be a cosmic coincidence (quite literally) or it may hint at an underlying explanation. \\

\subsection{Alternative Directions}

These difficulties are so major that they have motivated the exploration and development of alternatives to a strict cosmological constant or pure vacuum energy interpretation. These come in the form of either dynamical dark energy models where the gravitational sector is left essentially intact, and the goal is to find a more interesting dynamical mechanism that can describe dark energy, or a more profound modification of gravity in the infrared (\ie at large distances). Massive gravity occupies a distinctive position within the class of modified gravities in that it is the most theoretically constrained extension of Einstein's theory (other than a cosmological constant) requiring a minimal number of parameters. Endowing the graviton with a small mass which is tuned to be close to the current Hubble expansion rate modifies gravitational interactions on cosmological scales while preserving agreement with local tests through the nonlinear Vainshtein screening mechanism \cite{Vainshtein:1972sx} as explained below.

\subsection{How does massive gravity address the Cosmological Constant Problem?}

\paragraph{Infrared modification of gravity:}

Proposals to tackle the Old Cosmological Constant Problem typically focus on assuming that the Cosmological Constant is small and suggesting a mechanism to tame its quantum corrections. However, another possibility worth exploring is whether the Cosmological Constant is itself not small, in fact it could be as large as quantum corrections for Standard Model fields want it to be, possibly as large as the TeV scale or even larger; hence it does not suffer from any naturalness issues, but rather it is {\it perceived} to be so small because it gravitates weakly and hence as a much smaller effect of the curvature of the Universe compared to what would have been expected in GR,
 \cite{Dvali:2007kt,Arkani-Hamed:2002ukf,deRham:2007xp}. An explicit construction based on extra dimensions was proposed in \cite{deRham:2009wb} whereby an arbitrarily large cosmological constant gravitates so weakly that it leads to a four-dimensional Minkowski spacetime. 

To understand the idea behind this mechanism, first consider symbolically promoting Einstein's equations to a theory where the Newton's constant effectively depends on the momentum of the source:
\ba
G\mn=8\pi G_N T\mn\quad \Rightarrow \quad G\mn = 8\pi \int \d^4 k\  G_N^{\rm eff}(k)T^{(k)}\mn \,.
\ea
For GR, at the linear level about flat spacetime, we symbolically have 
 \ba
 \Box h\mn\sim G_N T\mn \quad \Rightarrow \quad h\mn\sim \frac{G_N}{\Box}T\mn\,.
 \ea
 Now in a theory of  massive gravity, where the graviton has a mass $m$, 
 \ba
\( \Box -m^2\) h\mn\sim G_N T\mn \quad \Rightarrow \quad h\mn\sim \frac{G_N}{\Box-m^2}T\mn\sim \frac{G_N^{\rm eff}}{\Box}T\mn\,,
 \ea
 with
 \ba
 G_N^{\rm eff}(k)=G_N\frac{k^2}{k^2+m^2}\ \to \ \begin{cases}
G_N & {\rm when } \ k\gg m\\
0 & {\rm when } \ k\ll m
 \end{cases}\,,
 \ea
 and so linearly, massive gravity can be interpreted as a theory of gravity where the effective Newton constant is promoted to a high-pass filter in the sense that at high frequency $G_N^{\rm eff} \to G_N$ and we recover GR for any source that has a large frequency as compared to the graviton mass. However, at small frequencies, the effective Newton constant essentially switches off, meaning that sources similar to a  cosmological constant (which effectively have zero or very small frequency) would gravitate very weakly (or not at all). 

To tackle the cosmological constant in this way, gravity must be weaker in the infrared and hence effectively massive (above we consider the case of a hard mass graviton; generalisations of this idea include resonances and soft masses, but for all intense and purposes, any promotion of the Newton constant such that it switches off at low frequencies must correspond effectively to a theory of massive gravity).

In such scenarios, if the vacuum energy had been a constant for an infinite amount of time, it would not gravitate at all, which explains why such models lead to exactly flat Minkowski solutions in the presence of a cosmological constant. However, in practice this mechanism is a causal process that starts at the same time as the Universe itself. For short periods of time as compared to the inverse graviton mass, the effect of the graviton mass is small, and the cosmological constant leads to the same outcome as in GR. It is only when the Universe is old enough (on a scale set by the inverse graviton mass) that the relaxation of the infrared cutoff takes effect, leading to a weakening and effective screening of the cosmological constant (in a linear theory, one would anticipate a Yukawa-type suppression). Consequently, inflation and the physics of the Early Universe proceed essentially unaffected by the graviton mass, and modifications to gravity become significant only at late cosmological times.

\paragraph{Mass tuning and technical naturalness:}
For this avenue to work, one does require the introduction of a new tuned parameter (the graviton mass). If massive gravity is to have an effect on cosmological scales, one would expect the graviton mass $m$ to be of the order of the Hubble parameter today $m\sim h_0\sim 10^{-33}$eV. In terms of tuning, this is of the same order of magnitude as the original cosmological constant problem $\mpl^{-2}\Lambda_{\rm C.C.}\sim H_0^{2}$. However, the key difference between tuning the cosmological constant to such small values and tuning the graviton mass is that the former receives large quantum corrections whereas the latter is protected against quantum corrections and hence is a technically natural tuning \cite{deRham:2012ew,deRham:2013qqa}. 
The key difference is that, since supersymmetry has to be broken already at about the TeV scale (or higher), there is no remaining symmetry that can protect the cosmological constant from becoming large. Indeed, a nonzero cosmological constant leads to a de Sitter spacetime, which is a maximally symmetric space, analogous to how Minkowski spacetime is maximally symmetric when the cosmological constant vanishes. So  $\Lambda_{\rm C.C.}=0$ leads to as much symmetry as $\Lambda_{\rm C.C.}\ne 0$ and there are no natural arguments to protect $\Lambda_{\rm C.C.}$ close to its vanishing value.  For the graviton mass, however, the crucial distinction lies in diffeomorphism invariance. In the massless case, \ie for GR—the theory is clearly protected by (non-linear) diffeomorphism invariance, implying that if the graviton mass vanishes, so do its quantum corrections. Equivalently, the quantum corrections $\delta m^2$ to the graviton mass scale proportionally to the mass itself, $\delta m^2 \sim m^2$. Indeed, it was demonstrated in \cite{deRham:2012ew,deRham:2013qqa} that the quantum corrections to the graviton mass behave as
\ba
\delta m^2\lesssim m^2\(\frac{m}{\mpl}\)^{1/2}\,,
\ea
so tuning the graviton mass to small values is a technically natural tuning that is protected from quantum corrections. 

This key insight therefore opens up the possibility of weakening gravity in the infrared in a way that remains technically stable. Within this framework, massive gravity provides a concrete and theoretically well-controlled model in which vacuum energy can interact with gravity differently on very large scales. It stands as an illustrative example of a gravitational EFT that retains many features of GR, while simultaneously displaying some highly distinctive differences \cite{deRham:2014zqa,deRham:2023byw}.

\vspace{0.3cm}
\noindent \textbf{\color{color1}Bottom Line\footnote{Format suggested by FQxI.}:} The (old) cosmological constant problem lies in the enormous discrepancy between the vacuum energy density required to account for the observed late-time accelerated expansion of the Universe and the value expected from QFT. In standard QFT, a particle of mass $m$ is generically expected to contribute to the vacuum energy with an amount of order $m^4$, leading, to a mismatch of roughly 56 orders of magnitude from Standard Model fields alone. While most proposed resolutions  either suppress the vacuum energy itself or make it  irrelevant, the original motivation for massive gravity was instead to pursue a technically natural solution by modifying gravity in the infrared. In this picture, gravity becomes weaker on sufficiently large scales, effectively acting as a high-pass filter that suppresses the gravitational response to vacuum energy while leaving early Universe and local gravitational phenomena relatively unaffected.

\section{(Beyond) GR as an Effective Field Theory}

The cosmological constant problem is one of the most vexing challenges of modern theoretical physics as it lies at the interface of the two pillars of the fundamental laws of Nature: the quantum realm of particle physics and General Relativity. The setting of the quantum vacuum of the cosmological problem may at first sight appear reminiscent of another well-known challenge of modern physics, namely that of \emph{Quantum Gravity}, which relates to the completion of General Relativity as a quantum field theory at energy scales close to the Planck scale. With that in mind, one may naturally be compelled to seek the resolution of the cosmological constant problem within that of the ultimate completion of quantum gravity (\eg string theory \cite{Bousso:2000xa,Witten:2000zk}, loop quantum gravity \cite{Smolin:2002sz}, causal sets \cite{Ahmed:2002mj}). 

However, both challenges lie on diametrically opposed energy scales. The realm of quantum gravity lies within the remit of the Planck scale \ien, in the deep UV (ultraviolet). However, the cosmological constant problem or vacuum energy probes the weakest possible energy scales in the Universe, sitting some $30$ orders of magnitude lower in the deep infrared regime. Rather than seeking a resolution of the Cosmological Constant problem within the realm of Quantum Gravity (\ie by embedding GR within a deeper UV completion), in these notes, we shall explore a resolution directly within the infrared modification of GR. 

Implicit in most of modern physics is the realisation that gravity is to be understood as a quantum Effective Field Theory (EFT), where the GR Einstein-Hilbert term is only the first term in an infinite low-energy expansion, so we shall review this quickly before moving into its generalisation.

\subsection{GR as an EFT}
\label{sec:GravityEFT}

To derive the EFT of gravity, we should first specify the low-energy field content and the symmetries. When treating GR as an EFT, we consider the field content to be the massless graviton as well as all other light (matter) fields that need to be considered to describe the relevant dynamics (\eg all the fields of the Standard Model or Particle Physics, Dark Matter, Dark Energy, potential other hidden sectors,$\dots$). GR is built on (nonlinear) diffeomorphism invariance (general covariance).
One crucial aspect one needs to take care of when dealing with the EFT of gravity is the diffeomorphism symmetry that GR relies on has one feature which is very different from Yang-Mills type gauge symmetries: There exist no \emph{local} gauge invariant operators in gravity.

Diffeomorphism invariance demands the action (in $D=4$ dimensions for concreteness) takes the form
\ba
S = \int \d^4 x \sqrt{-g} {\cal L}(x) \, ,
\ea
where ${\cal L}(x)$ is a scalar. With this in mind, the EFT of gravity is naturally organised as an expansion in derivatives that are increasingly irrelevant operators. Up to fourth  order in derivatives, the unique scalar operators that can be included in the EFT of gravity are 
\ba
\label{eq:GREFT}
{\cal L} =  \frac{\mpl^{D-2}}{2} (R-2\Lambda) + \left[  c_0 \Box R+c_1 R^2 +c_2 R_{\mu\nu}^2+c_3 R_{\mu\nu\rho \sigma}^2  \right] + {\cal O}(R^3)\, , 
\ea
where $\Lambda$ is the Cosmological Constant we already introduced earlier and the parameters $c_n$ are dimensionless Wilsonian coefficients which capture information about the UV physics (we may for instance expect them to depend primarily on the masses of the lightest high-spin states that enter the UV physics). Here, the cosmological constant is introduced in the normalisation that Einstein used, but it is perhaps more natural to introduce it as a vacuum energy $\rho_{\Lambda} =\mpl^{2} \Lambda $ so that
\ba
{\cal L} = -\rho_{\Lambda}+ \frac{\mpl^{2}}{2} R + \left[  c_0 \Box R+c_1 R^2 +c_2 R_{\mu\nu}^2+c_3 R_{\mu\nu\rho \sigma}^2  \right] + {\cal O}(R^3)\, .
\ea
From this perspective, the cosmological constant, or really $\rho_{\Lambda}$, is the most relevant operator in the standard theory of GR
and is clearly of dimension $4$.
The $R$ operator is the unique dimension $2$ scalar-invariant built out of the metric and is nothing other than the standard GR Einstein-Hilbert term. 
Phrased in this way, the real challenge related to the cosmological constant problem is the fact that the Einstein-Hilbert term is not the most relevant operator in the EFT of gravity, the Cosmological Constant is!
Independently of the number of dimensions, both the Cosmological Constant and the Einstein--Hilbert terms are the unique scalar operators at their respective order \cite{Wald:1986bj}.

The fact that the Cosmological Constant and the Einstein--Hilbert terms are the unique ones at their respective order is an accident. Beyond this order, the number of operators increases significantly. Up to dimension-12 and up to field redefinitions, the independent local and parity-even operators entering the EFT of gravity are given by  \cite{deRham:2022gfe} 
\ba
\label{eq:EFT1}
\L=-\rho_{\Lambda}+\frac{\mpl^{2}}{2}R+\frac{c_3}{M^2} R_{\mu\nu\rho \sigma}^2+\frac{c_1}{M^4} \mathcal{C}^2+\frac{c_2}{M^4}\Tilde{\mathcal{C}}^2+\frac{e_1}{M^6}[\mathcal{F}]\mathcal{C}+\frac{e_2}{M^6}[\tilde{\mathcal{F}}]\tilde{\mathcal{C}} \\
+ \frac{f_1}{M^8}[\mathcal{F}]^2+\frac{f_2}{M^8}[\tilde{\mathcal{F}}]^2+
\frac{g_1}{M^8}[\mathcal{F}^2]+\frac{g_2}{M^8}[\tilde{\mathcal{F}}^2]+\frac{j_1}{M^8} \mathcal J
\,,\nn
\ea
where we have again focused on $D=4$ dimensions for definiteness and defined
\ba
	R^3= R_{\alpha\beta\rho\sigma}R^{\rho\sigma}_{\phantom{\alpha\beta}\mu\nu}R^{\mu\nu\alpha\beta}\,,\quad\mathcal{C}=R_{\mu\nu\rho\sigma}R^{\mu\nu\rho\sigma}\,,\quad\Tilde{\mathcal{C}}=\frac12R^{\alpha\beta\mu\nu}\epsilon_{\mu\nu\rho\sigma}R^{\rho\sigma}_{\phantom{\rho\sigma}\alpha\beta}\,, \nn \\
	\mathcal{F}_{\alpha\beta}= \nabla_{\alpha}R_{\mu\nu\rho\sigma}\nabla_{\beta}R^{\mu\nu\rho\sigma}\,,\quad\tilde{\mathcal{F}}_{\alpha\beta}=\frac12 \nabla_{\alpha}R^{\gamma\delta \mu\nu}\nabla_\beta\left(\epsilon_{\mu\nu\rho\sigma}R^{\rho\sigma}_{\phantom{\rho\sigma}\gamma\delta}\right)\,,\qquad
\ea
and
$$
\mathcal J = \nabla^\mu\nabla^\nu R^{\alpha\beta\gamma\delta}\nabla^\varepsilon R^{\chi}_{\phantom{\chi}\zeta\delta \gamma}\nabla^{\zeta}R_{\varepsilon\alpha \iota\beta}R^{\iota}_{\phantom{\iota}\nu\mu\chi}\,.
$$
In addition to these parity-even  operators, the EFT of gravity to that order also  includes  5 parity-odd operators $\mathcal{C} \Tilde{\mathcal{C}}, [\mathcal{F}]\Tilde{\mathcal{C}}, [\mathcal{F}] [\Tilde{\mathcal{F}}], [\mathcal{F}\Tilde{\mathcal{F}}]$ and $\Tilde{\mathcal{J}}$, which is dual to  $\mathcal{J}$.  

\subsection{Beyond GR as an EFT}

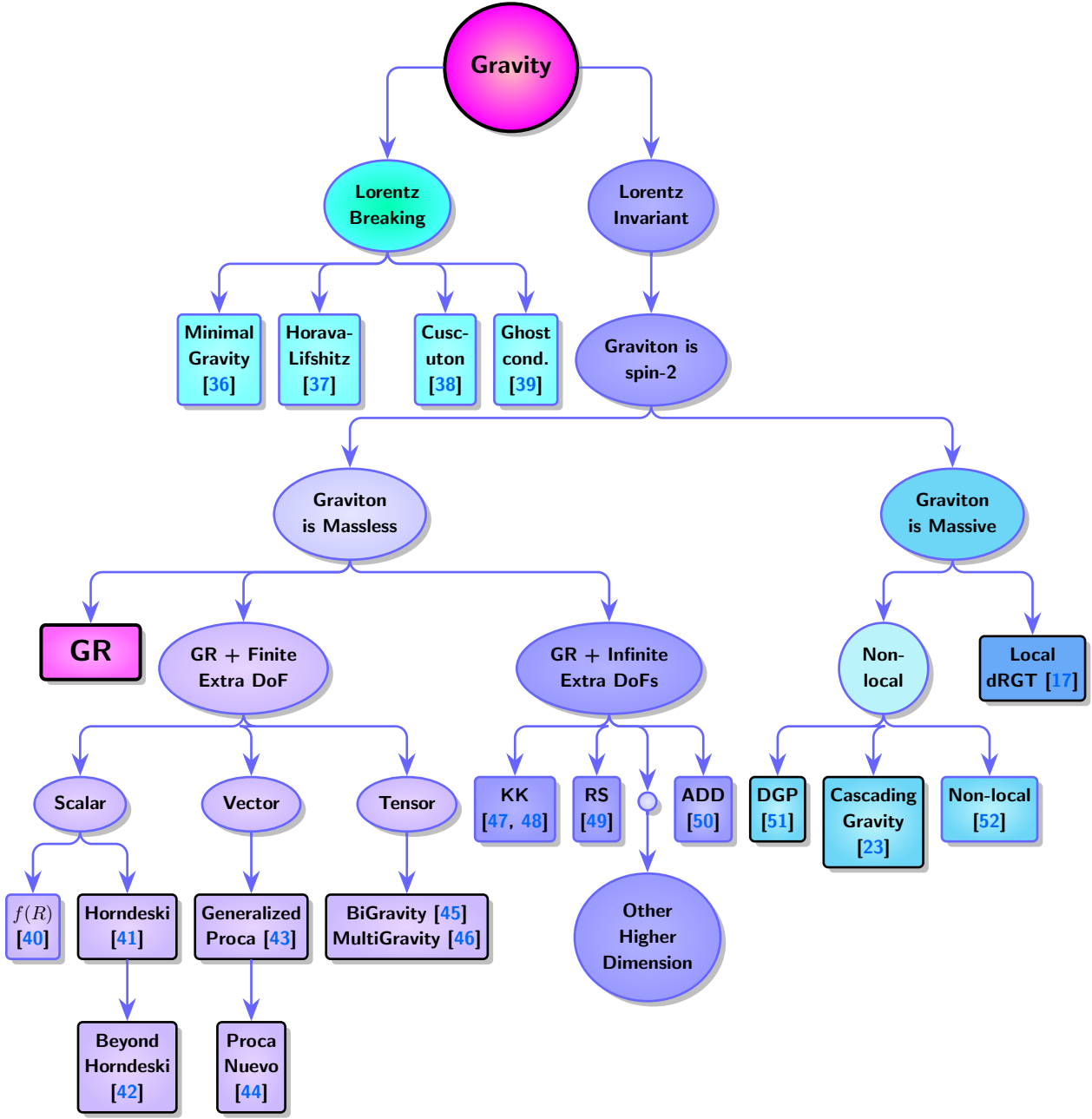
\begin{figure}[h!]


\pgfkeys{/forest,
  rect/.append style   = {rectangle, rounded corners = 2pt,
                         inner color = col6in, outer color = col6out},
  ellip/.append style  = {ellipse, inner color = col5in,
                          outer color = col5out},
  orect/.append style  = {rect, font = \sffamily\bfseries\LARGE,
                         text width = 325pt, text centered,
                         minimum height = 10pt, outer color = col7out,
                         inner color=col7in},
  oellip/.append style = {ellip, inner color = col8in, outer color = col8out,
                          font = \sffamily\bfseries\large, text centered}}
\hspace{-0.8cm}\begin{forest}
  for tree={
      scale=0.75,
      font=\sffamily\bfseries,
      line width=1pt,
      draw=linecol,
      ellip,
      align=center,
      child anchor=north,
      parent anchor=south,
      drop shadow,
      l sep+=12.5pt,
      edge path={
        \noexpand\path[color=linecol, rounded corners=5pt,
          >={Stealth[length=10pt]}, line width=1pt, ->, \forestoption{edge}]
          (!u.parent anchor) -- +(0,-5pt) -|
          (.child anchor)\forestoption{edge label};
        },
      where level={3}{tier=tier3}{},
      where level={0}{l sep-=15pt}{},
      where level={1}{
        if n={1}{
          edge path={
            \noexpand\path[color=linecol, rounded corners=5pt,
              >={Stealth[length=10pt]}, line width=1pt, ->,
              \forestoption{edge}]
              (!u.west) -| (.child anchor)\forestoption{edge label};
            },
        }{
          edge path={
            \noexpand\path[color=linecol, rounded corners=5pt,
              >={Stealth[length=10pt]}, line width=1pt, ->,
              \forestoption{edge}]
              (!u.east) -| (.child anchor)\forestoption{edge label};
            },
        }
      }{},
  }
  [\\ {\Large Gravity} \\ $\phantom{.}$, inner color=pink,
  outer color=magenta, line width=1.5pt, color=black
    [Lorentz\\Breaking, inner color=cyan!70!green, outer color=cyan!70!white
      [Minimal \\Gravity\\ \cite{DeFelice:2015hla}, rect, inner color=cyan!30!white, outer color=cyan!50!white
       ]
      [Horava-\\ Lifshitz \\ \cite{Horava:2009uw}, rect, inner color=cyan!30!white, outer color=cyan!50!white]
          [, phantom, calign with current]
       [Cusc- \\ uton \\ \cite{Afshordi:2006ad}, rect, inner color=cyan!30!white, outer color=cyan!50!white]
       [Ghost\\ cond. \\  \cite{Arkani-Hamed:2003pdi} , rect, inner color=cyan!30!white, outer color=cyan!50!white]
    ]
    [, phantom, calign with current]
    [, phantom, calign with current]
    [Lorentz\\Invariant, inner color=col3in, outer color=col3out
     [Graviton is\\ spin-2, inner color=col3in, outer color=col3out
        [Graviton \\ is Massless,  name=sse1
            [ GR , rect, font = \sffamily\bfseries\LARGE,
                         text width = 50pt, text centered,
                         minimum height = 10pt, outer color = magenta!60!white,
                         inner color=magenta!30!white, line width=1.5pt, color=black
            ]
            [GR + Finite  \\  Extra DoF, inner color=col9in, outer color=col9out
                [Scalar, inner color=col9in, outer color=col9out
                    [$f(R)$ \\ \cite{Carroll:2003wy}, rect, inner color=col9in, outer color=col9out]
                    [Horndeski \\ \cite{Horndeski:1974wa}, rect,  inner color=col9in, outer color=col9out, color=black  
                        [Beyond \\ Horndeski \\ \cite{Gleyzes:2014dya}, rect, inner color=col9in, outer color=col9out, color=black
                        ]
                    ]
                ]
                [Vector, inner color=col9in, outer color=col9out
                    [Generalized \\ Proca \cite{Heisenberg:2017mzp}, rect,  inner color=col9in, outer color=col9out, color=black 
                    [Proca \\ Nuevo \\ \cite{deRham:2020yet}, rect, inner color=col9in, outer color=col9out, color=black
                        ]
                ]]
                [Tensor, inner color=col9in, outer color=col9out
                    [BiGravity \cite{Hassan:2011zd}\\ MultiGravity \cite{Hinterbichler:2012cn}, rect, inner color=col9in, outer color=col9out, color=black
                    ]
                ]
            ]
           [GR + Infinite \\ Extra DoFs, inner color=col3in, outer color=col3out
                [KK \\ \cite{Kaluza:1921tu,Klein:1926tv} , rect, inner color=col3in, outer color=col3out]
                [RS \\ \cite{Randall:1999vf} , rect, inner color=col3in, outer color=col3out]
                [[Other\\ Higher  \\ Dimension, inner color=col3in, outer color=col3out]]
                [ADD \\ \cite{Arkani-Hamed:1998jmv}, inner color=col3in , rect, outer color=col3out]
            ]
        ]
        [Graviton \\ is Massive, inner color=NLout, outer color=NLout
            [Non-\\ $\phantom{.}$local$\phantom{.}$, inner color = NLin, outer color = NLin
                [DGP \\ \cite{Dvali:2000hr}, rect, inner color = NLin, outer color = NLout,  color=black]
                [Cascading\\  Gravity \\  \cite{deRham:2007xp}, rect, inner color = NLin, outer color = NLout,  color=black]
                [Non-local \\ \cite{Jaccard:2013gla}, rect, inner color = NLin, outer color = NLout]
            ]
            [, phantom, calign with current]
            [Local \\ dRGT \cite{deRham:2010kj}, rect, inner color=MGin, outer color=MGin,  color=black
            ]
        ]
    ]
    ]
  ]
  \begin{scope}[color = linecol, rounded corners = 5pt,
    >={Stealth[length=10pt]}, line width=1pt, ->]
  \end{scope}
\end{forest}


  \caption{Classification of Models of Modified Gravity. Adapted from~\cite{deRham:2019wjj} and \cite{deRham:2023byw}. Bold boxes include Galileon degrees of freedom.}
  \label{fig:MGmodels2}
  \end{figure}

To go beyond GR—especially when treated as an EFT—a natural strategy is to alter either the low-energy field content or the underlying symmetry on which it is based.  
Another possibility is to change how gravity couples to other dynamical fields (the ‘matter fields’).
In many situations these different perspectives feed into each other, since the breaking of local symmetries necessarily leads to the presence of additional degrees of freedom with non-minimal couplings to matter fields \cite{Clifton:2011jh,deRham:2019wjj}. The classification of the various modifications of gravity depending on the symmetry being preserved, the field content, and whether locality is preserved is indicated in Fig.~\ref{fig:MGmodels2}. Different models beyond GR address various challenges and these are summarised in  Table~\ref{Table:MG_motivations}. 

\paragraph{Galileons:} In Fig.~\ref{fig:MGmodels2}, many models feature what are commonly referred to as  `Galileon' degrees of freedom \cite{Nicolis:2008in}. These typically arise in models of modified gravity where the graviton has additional polarisation. In certain regimes, the behavior of these modes can be conveniently captured by an effective scalar degree of freedom $\pi$, which possesses a non-trivial space-dependent global symmetry (that is, it remains invariant under $\pi \to \pi + c + v_\mu x^\mu$). Despite involving higher-derivative interactions, this scalar remains free from ghost-like instabilities.

\begin{table}[h!]
\begin{center}
\begin{tabular}{|c|c|c||c||c|c|c|c|c|c|c|c|c|}
\hline
& \tn{IR behaviour} & \tn {UV behaviour} & \tn{Lorent inv. Vac.\ } & \tn{UV finite}& \tn{Unification}& \tn{Hierarchy}& \tn{Dark Matter} &\tn{Dark Energy} & \tn{CC Problem} &
\tn{Early Universe}& \tn{Cosmo Tensions\ \ } & \tn{Solar Anomalies\ \ }
 \\
 \hline
 %
KK & \tt{GR} & {\color{violet} \tt{Mod}} & \Checkmark    &  & \Checkmark    &   & \Checkmark   &  & & &   & \\ \hline
RS  & \tt{GR} &  {\color{violet} \tt{Mod}}  & \Checkmark    & & \Checkmark & \Checkmark & \Checkmark & & & & &   \\ \hline
ADD & \tt{GR} &  {\color{violet} \tt{Mod}} & \Checkmark    &  & & \Checkmark &  & \Checkmark & &  & &   \\ \hline
SLED & \tt{GR}  & {\color{violet} \tt{Mod}} & \Checkmark    &    &   &    \Checkmark & & \Checkmark & \Checkmark & &   &\\ \hline
Horava-Lifshitz & \tt{GR} &  {\color{violet} \tt{Mod}}& & \Checkmark   &  &     &   &  & & &   & \\ \hline
Stelle gravity & \tt{GR} &   {\color{violet} \tt{Mod}}& & \Checkmark  & &     &   &  & & &   & \\ \hline
Non-Local Gravity &\tt{GR?} &  {\color{violet} \tt{Mod}}& \Checkmark  & \Checkmark    & &     &   & \Checkmark & & &   & \\ \hline
\hline
MOND/TeVeS &  {\color{red} \tt{Mod}} &  \tt{GR} & {\color{gray}\Checkmark}& &  &  & \Checkmark         & & & &   & \Checkmark \\ \hline
DGP/Cascading &{\color{red} \tt{Mod}} &  \tt{GR} & \Checkmark&    &  & & & \Checkmark    &  \Checkmark & & &   \\ \hline
Massive Gravity & {\color{red} \tt{Mod}} &  \tt{GR?}& \Checkmark&   &  &   &  & \Checkmark   & \Checkmark & \Checkmark & \Checkmark  &\Checkmark \\ \hline
Minimal Gravity &{\color{red} \tt{Mod}} &  \tt{GR?}& \Checkmark& &  &     &   & \Checkmark  & & \Checkmark& \Checkmark  & \\ \hline
\hline
Cuscuton &  {\color{red} \tt{Mod}$_+$}& \tt{Mod} & &  &  &   &   &  \Checkmark & & &   &  \\ \hline
Ghost Condensate &{\color{red} \tt{Mod}$_+$} &  \tt{GR?}& & &  &     &   & \Checkmark & & &   & \\ \hline
$f(R)$ & \tt{GR}$_+$  &  \tt{Mod}& \Checkmark& &  &     &   &  \Checkmark  & & &   & \\ \hline
Horndeski & \tt{GR}$_+$ &  \tt{GR?}&\Checkmark&  &  &     &   & \Checkmark & &\Checkmark &\Checkmark  & \\ \hline
DHOST & \tt{GR}$_+$  &  \tt{GR?}& \Checkmark& &  &     &   & \Checkmark  & &\Checkmark &  \Checkmark & \\ \hline
Bi/Multi-Gravity & \tt{GR}$_+$  &  \tt{GR?}& \Checkmark&    &  &  &  \Checkmark & \Checkmark   & & &  \Checkmark &\\ \hline
GP/PN & \tt{GR}$_+$  & \tt{GR?}& \Checkmark& &  &     &   & \Checkmark & & \Checkmark & \Checkmark  & \\ \hline
\end{tabular}
\end{center}
 \caption{\label{Table:MG_motivations}(Non-exhaustive) examples of theories beyond GR and challenges they may tackle (indicated with a checkmark, which does not necessarily imply a resolution but rather a regime in which these challenges may be tackled/ potentially addressed). Most models behave like `{\tt{GR}}' either in the UV or in the IR, while `{\tt{Mod}}' indicates in which regime they are modified. Some models behave as in GR in the IR with the addition of (potentially screened) degrees of freedom those are  indicated as {\tt{GR}}$_+$. Models  {\tt{Mod}}$_+$ indicate a non-trivial vacuum for the additional degrees of freedom. Grey checkmarks indicate that it depends on the model or implementation. Adapted from~\cite{deRham:2023byw}. 
 SLED is introduced in \cite{Aghababaie:2003wz},
 Stelle Gravity in \cite{Stelle:1976gc} and
 MOND in \cite{Bekenstein:1984tv,Milgrom:1998aj}.
The other models are referred to in Fig.~\ref{fig:MGmodels2} or in the text.}
\end{table}

\paragraph{Screening:} 
As can be seen in the previous figure and table, many models of modified gravity include additional degrees of freedom. When those couple to the Standard Model fields, the phenomenological viability of these models requires the presence of screening mechanisms that naturally (and often dynamically) weaken the force propagated by these fields, at least in environments where GR has been well tested. The chameleon is a well-known example through which a scalar degree of freedom can become very massive in regions with large energy density \cite{Khoury:2003aq} and is realised in $f(R)$. Another kind of screening mechanism effectively makes the additional degrees of freedom weakly coupled to Standard Model fields as is the case of the symmetron  \cite{Hinterbichler:2010es}.
Finally, the Vainshtein mechanism relies on nonlinear derivative interactions (such as Galileon-type terms) to redress the effective kinetic term of the extra degree of freedom, thereby suppressing its dynamics in non-trivial configurations \cite{Vainshtein:1972sx,Babichev:2013usa}. These mechanisms play an important role in reconciling theories of  gravity that include additional degrees of freedom in the IR, with precise tests of gravity in the solar-system and the laboratory, while still allowing for potentially observable effects on cosmological scales. 

Within this broad landscape, massive gravity occupies a particularly distinctive position. Rather than introducing new light scalar degrees of freedom in their own right, or changing the coupling to matter fields, massive gravity directly alters the gravitational degree of freedom, namely the properties of the graviton by endowing the graviton with a small but finite mass. As a result, the propagating degrees of freedom differ qualitatively from those of GR, with the massless spin--2 field replaced by a massive spin--2 representation carrying additional polarisations. Unlike models of modified gravity that rely on extra dimensions, massive gravity remains intrinsically four-dimensional and local, although the modifications of gravity it induces on large distances are similar to some models of gravity with large extra dimensions \`a la DGP \cite{Dvali:2002fz} or Cascading Gravity \cite{deRham:2007xp}. In fact, any model of gravity that genuinely modifies the dynamics of gravitational degrees of freedom in the infrared can effectively be recast as a (local or non-local) model of gravity where gravity acquires a mass or resonance \cite{Dvali:2007kt}. Exploring a local theory of gravity where the graviton remains a spin--2 degree of freedom but endowed with a small mass hence  
provides a self-contained framework in which many models of modified gravity can be analysed while using standard effective field theory tools.

\subsection{Spin--2 EFT}

From a conceptual point of view, the idea behind massive gravity is very natural. The weak force is Nature's concrete realisation of force carrier particles endowed with a mass weakening the strength of these forces in their infrared (in this case on energy scales below the EW mass, or on distance scales longer than the W and Z boson Compton wavelengths). Whether or not a similar phenomenon may occur for gravity is hence a profound question to be explored in its own right, independently of the cosmological constant problem or other challenges such a feature may potentially address. Just as endowing spin--1 fields with a small mass leads to additional polarisations (degrees of freedom), the same happens for massive gravity as first identified by Fierz and Pauli in 1939 \cite{Fierz:1939ix}.

The notions of mass and spin presuppose Lorentz invariance (they are defined by the representation theory of the Poincar\'e group) and so technically speaking the very notion of mass (just as that of spin of a fundamental particle) requires a fundamental Minkowski metric. It is therefore natural to start by exploring the structure of massive gravity perturbatively about Minkowski, in the same way as all particles of the Standard Model are defined via their Lorentz representation. 
 
To start with, we may therefore write the metric as a deviation from Minkowski
\begin{equation}
g_{\mu\nu} = \eta_{\mu\nu} + \bar h_{\mu\nu} \,.
\end{equation}
In principle, this representation is exact and $\bar h_{\mu\nu}$ does not need to correspond to a small deviation from Minkowski spacetime. However, to start with, it is instructive to limit ourselves to that consideration. 

\paragraph{Kinetic term:} As we shall see, the presence of the mass term breaks diffeomorphism invariance, hence opening up many new possibilities for its kinetic term, which is in principle no longer restrained to take a specific structure and hence no longer unique. This possibility was thoroughly explored in the literature \cite{Comelli:2012vz,Hinterbichler:2013eza,Comelli:2013txa,deRham:2013tfa,Goon:2014paa,Noller:2014ioa,Gao:2014jja,deRham:2015rxa,deRham:2015cha,Matas:2016qjj,Bonifacio:2017iry,Alberte:2019lnd} with the conclusion that only the standard Einstein--Hilbert kinetic term is unitary as the leading term in a derivative expansion EFT. Interestingly, in the context of a theory that does not couple to other dynamical fields, an alternative kinetic interaction was proposed in \cite{Hinterbichler:2013eza}. At first sight, this interaction appears reminiscent of the would-be Lovelock terms that arise in higher dimensions. However, no consistent completion of this operator is known that would allow the theory to couple to other dynamical fields, \ie elevate it to a genuine gravitational theory, without introducing a ghost at an unacceptably low scale \cite{deRham:2013tfa}.
This is a remarkable result in its own right, as it indicates that as a kinetic term, the Einstein one is uniquely determined without needing to invoke diffeomorphism invariance. In other words, there is no need to rely on Einstein's pillar of GR (covariance) to infer that GR is the unique theory of a massless spin--2 field.

Working in terms of the canonically normalised field $\bar h\mn=\tilde h\mn/\mpl$, the kinetic term takes the form
\ba
S_{\rm EH}&=&\frac{\mpl^2}{2}\int \d^4 x \sqrt{-g}R \\
&=& \int \d ^4x \(-\frac 14 \tilde h^{\mu\nu}\hat{\E}^{\alpha \beta}\mn \tilde h\ab + \frac{1}{\mpl}\tilde h (\p \tilde h)^2 + \cdots\)\,,
\ea
where $\hat{\E}^{\alpha \beta}\mn \tilde h\ab=-\frac 12 \Box \tilde h\mn+\cdots$ is the Lichnerowicz operator, and the ellipses represent an infinite number of nonlinear terms, precisely tuned so as to be fully covariant. Crucially, we recover the linearised structure in the limit $\mpl\to \infty$.

\paragraph{Fierz-Pauli mass term:} 
Having identified the Einstein--Hilbert term as the unique dimension--2 kinetic term for any spin--2 field, (both massless and massive), the next step in constructing the EFT of a massive spin--2 field is to identify its proper mass structure. In the same way that the kinetic term cannot be chosen as an arbitrary function of the field (for example, $h^{\mu\nu}\Box h\mn$ by itself would not provide a consistent kinetic term), the mass term is similarly constrained and must also appear in a particular form. As shown by Fierz and Pauli, already at the linear level,  Locality, Poincar\'e invariance, and Unitarity restrict the form of admissible quadratic interactions, leading to a unique structure up to an overall normalisation \cite{Fierz:1939ix},
\begin{equation}
\mathcal{U}_{\mathrm{FP}} = m^2\mpl^2 \(\bar h_{\mu\nu} \bar h^{\mu\nu} - \bar h^2\) ,
\end{equation}
where $\bar h = \bar h^\mu{}_\mu$ denotes the trace of $\bar h_{\mu\nu}$.
When combined with the Einstein–Hilbert kinetic term, this mass term is the only one (at that order) that guarantees the \emph{linearised equations of motion} propagate precisely five physical degrees of freedom. This is the correct number of degrees of freedom for a massive spin--2 field in $D=4$ dimensions, in agreement with the representation theory of the Poincar\'e group for a massive spin--2 particle. From that perspective, the Fierz--Pauli mass term is not an arbitrary choice, but a necessary consistency condition required to eliminate unphysical (non-unitary or ghost-like) excitations, at that order \cite{Fierz:1939ix}.

\paragraph{Interacting EFT:} In its own right, in the absence of any couplings (neither to other dynamical fields nor nonlinear self-interactions), together with the linearised Einstein--Hilbert term, the Fierz-Pauli mass term leads to a perfectly consistent theory of a free massive spin--2 field. However, already for a massless spin--2 field (\ie GR), it is well-known that consistently coupling the massless spin--2 field to other dynamical fields requires promoting the linearised Einstein--Hilbert term to the full curvature term, yielding an infinite number of nonlinear interactions. It should therefore come as no surprise that the same should be done for the Fierz-Pauli mass term when promoting the theory to an interacting one. The resulting nonlinear theory can be written schematically as
\begin{equation}
\label{eq:massive}
S_{\text{Massive Gravity}} = \frac{M_{\mathrm{Pl}}^2}{2}\int \d^4x \sqrt{-g}\, 
\left( R - m^2\,  \mathcal{U}[g,\eta] \right)+\int \d^4x \sqrt{-g}\, \L_{\rm matter}[g\mn \psi_i]\,,
\end{equation}
where $m$ denotes the graviton mass, $\eta_{\mu\nu}$ is the reference metric, and $\mathcal{U}$ is a scalar potential constructed from contractions of $g_{\mu\nu}$ and the reference metric $\eta_{\mu\nu}$. For completeness, we have also introduced the standard matter Lagrangian $\L_{\rm matter}$ in terms of all dynamical fields $\psi_i$. Note that the theory is implicitly defined with the boundary terms derived in \cite{Gabadadze:2018bpf}. 
As was already the case in the Fierz-Pauli action, the introduction of the mass term explicitly breaks diffeomorphism invariance. As a result, the constraint structure of the theory is modified as compared to GR, and the theory involves additional degrees of freedom (just as was the case when endowing a spin-1 field with a mass term). Nevertheless, even without appealing to any symmetry-based guiding principle, the uniqueness argument—which demands that the kinetic term match that of GR—extends equally to the matter coupling~\cite{deRham:2013tfa,Yamashita:2014fga,deRham:2014naa,deRham:2014fha,deRham:2015rxa,deRham:2015cha,Matas:2015qxa}.

The presence of the mass implies that waves of these fields do not propagate at the speed of light. In the case of a massive spin-1 field, the additional degree of freedom is directly linked with the longitudinal polarisation (excitations along the line of propagation, which are possible since the massive particles are timelike and not  constrained to propagate along null geodesics). In the case of massive spin--2 the connection between the additional polarisations and the longitudinal mode is much more subtle.
From a representation point of view, a massive spin--2 field in $D=4$ dimensions propagates five physical degrees of freedom: two helicity-2 modes that behave as tensors, two helicity-1 modes that behave as vector modes, and one helicity-0 mode that behaves as a scalar. The identification of these modes into Scalar, Vector, Tensor is not exact and only meaningful when taking a decoupling limit and upon identification of the residual global symmetry with the Lorentz one. 

An important breakthrough in the development of massive gravity was the exact identification of the full nonlinear Fierz-Pauli mass term, which maintains these five degrees of freedom and no other unphysical (non-unitary) excitations. 
This identification is now commonly referred to as de~Rham--Gabadadze--Tolley (dRGT) massive gravity and provides a ghost-free nonlinear completion of the Fierz--Pauli theory \cite{deRham:2010ik,deRham:2010kj}. Some precursors include the models by 
Maheshwari \cite{Maheshwari:1972mb}, Wess-Zumino \cite{Zumino:1970tu} and Ogievetsky-Polubarinov \cite{Ogievetsky:1965zcd}. These earlier constructions incorporated some of the ghost-free dRGT terms as well as other, ghostly ones, but they did not cover all possible terms, nor did they yet provide a clear understanding of the conditions required for the theories to be ghost free.

In dRGT massive gravity, the interaction potential is built from a specific tensor structure $\mathcal{K}^\mu{}_\nu$ involving the dynamical metric $g_{\mu\nu}$ and the reference metric $\eta_{\mu\nu}$ 
\begin{equation}
\mathcal{K}^\mu{}_\nu = \delta^\mu_\nu - \sqrt{g^{\mu\alpha} \eta_{\alpha\nu}} ,
\end{equation}
where the square root denotes the matrix square root, so $\K$ is implicitly defined by the relation
\ba
\K^\mu{}_\alpha \K^\alpha{}_\nu-2\K^\mu{}_\nu=-\delta^\mu{}_\nu+g^{\mu\alpha}\eta_{\alpha \nu}\,.
\ea
This quantity $\K$ is reminiscent of the extrinsic curvature that arises in extra dimensions \cite{Gabadadze:2009ja,deRham:2009rm,deRham:2010gu}, hence the name. With that building block in mind, the dRGT potential is given as
\begin{equation}
\label{eq:fullU}
\mathcal{U} =
\mathcal{U}_2[\mathcal{K}]
+ \alpha_3\, \mathcal{U}_3[\mathcal{K}]
+ \alpha_4 \, \mathcal{U}_4[\mathcal{K}] ,
\end{equation}
where $\mathcal{U}_n$ denote the symmetric polynomials of order $n$ built from $\K$ and the dimensionless parameters $\alpha_3$ and $\alpha_4$ label a two-parameter family of ghost-free theories,
\ba
\mathcal{U}_2[\mathcal{K}]&\sim& \E^{abcd}\E_{\alpha\beta cd}\K^\alpha{}_a \K^\beta{}_b \\
\mathcal{U}_3[\mathcal{K}]&\sim& \E^{abcd}\E_{\alpha\beta \gamma d}\K^\alpha{}_a \K^\beta{}_b \K^\gamma{}_c\\
\mathcal{U}_4[\mathcal{K}]&\sim& \E^{abcd}\E_{\alpha\beta \gamma \delta}\K^\alpha{}_a \K^\beta{}_b \K^\gamma{}_c \K^\delta{}_d\,,
\ea
where $\E^{abcd}$ represents the fully antisymmetric Levi-Cevita symbol.
This specific structure is uniquely fixed by unitarity requirements, or in other words, by the requirement that the theory does not propagate a ghost-like degree of freedom (the so-called Boulware--Deser ghost) \cite{deRham:2011rn,Hassan:2012qv}.

The absence of the Boulware--Deser ghost in dRGT massive gravity can be established using several complementary approaches, including Hamiltonian analyses formulated in \cite{deRham:2010kj} and extended to general situations in 
\cite{Hassan:2011hr,Hassan:2011ea,Kluson:2011qe,Golovnev:2011aa,Kluson:2011rt} and then to a generic reference metric $\eta\mn \to f\mn$ for an arbitrary reference metric \cite{Hassan:2011tf} or dynamical one, leading to bigravity \cite{Hassan:2011zd} as well as in the 
St\"uckelberg formulation \cite{deRham:2011rn,Hassan:2012qv,Kluson:2012cqq,Kluson:2012gz,Noller:2013yja}, 
the helicity one \cite{deRham:2011qq} and formal covariantisation of the \stu fields within and beyond the decoupling limit \cite{deRham:2010ik,deRham:2010kj,Mirbabayi:2011aa,Gao:2015xwa} as we shall see in Section~\ref{sec:DL}. While the precise technical implementations differ for each one, all of these analyses reveal the existence of a fully non-perturbative primary and secondary second class constraint which together ensures that the theory only propagates five degrees of freedom on generic backgrounds, as required for a massive spin--2 field in four dimensions. 

The construction of dRGT massive gravity represents a significant conceptual achievement. It demonstrates conclusively that ghost-free, Lorentz-invariant massive gravity theories do exist at the nonlinear level, resolving a long-standing obstruction to the formulation of a fully nonlinear interaction term that maintains the constraint in a completely non-perturbative way. At the same time, the highly constrained nature of the dRGT interaction potential, together with the fact that it allows only two independent dimensionless parameters ($\alpha_3$ and $\alpha_4$ in addition to the mass scale $m$), underscores how delicate it is to modify gravity in the infrared and underlines the rigidity of General Relativity, which still stands as the sole consistent kinetic term for gravity.

\vspace{0.3cm}
\noindent \textbf{\color{color1}Bottom Line:} GR is the leading term in an low-energy EFT sense of a massless spin-2 field.   In parallel, massive gravity directly modifies the graviton's propagator while 
remaining a local, four-dimensional and can be studied using the standard EFT framework. In dRGT massive gravity, a specific non-linear structure for the mass term can be identified that is free from the Boulware-Deser ghost and hence enjoys as high a strong coupling scale as possible, in a way that is parametrically dependent on the mass. 

\section{Propagating Degrees of Freedom}
\label{sec:DoF}

\subsection{St\"uckelberg Formalism}

A powerful way to analyse the theoretical and phenomenological consistency of massive gravity is to directly identify the dynamical degrees of freedom using the St\"uckelberg trick. The idea behind the \stu formulation is to explicitly restore diffeomorphism invariance by introducing additional fields that compensate for the explicit breaking induced by the graviton mass term. Nominally, this procedure amounts to a mere reformulation of the theory (through the introduction of new variables) that leaves its physical content unchanged, but yields an equivalent description in which the symmetry structure and constraint dynamics are explicit, often leading to a substantial simplification of the analysis \cite{ArkaniHamed:2002sp,Rubakov:2008nh}.

Concretely, diffeomorphism invariance can be restored by promoting the fixed reference metric $\eta_{\mu\nu}$ to a covariant object via the introduction of four {\bf scalar} \stu fields $\phi^a$. Note that at this stage, the index carried by the four \stu fields is not a Lorentz index but rather an internal space index spanning the four fields.
Thanks to these four fields, explicit 
diffeomorphism invariance is achieved by  replacing
\begin{equation}
\eta_{\mu\nu} \;\rightarrow\; \tilde \eta \mn = \partial_\mu \phi^a \partial_\nu \phi^b \eta_{ab},
\end{equation}
where $\eta_{ab}$ is now the Minkowski metric in the internal field space, so that in its ``St\"uckelbergised" version, the building block $\tilde\K\mupn$ is expressed as 
\ba
\tilde\K\mupn = \delta\mupn -\sqrt{g^{\mu\alpha} \tilde \eta_{\alpha \nu}}\,.
\ea
Crucially, under spacetime diffeomorphisms, the four \stu scalar fields $\phi^a$ transform as {\bf scalars}, ensuring that the quantity $\tilde \eta_{ab}$ transforms as a rank-two tensor. This directly implies that $\tilde \K\mupn$ is a rank $(1,1)$-tensor under diffeomorphisms and so the dRGT potential given by
\ba
\tilde \U=\tilde \U[g\mn, \phi^a]=\U_2[\tilde \K]+\alpha_3\, \U_3[\tilde \K]+\alpha_4\,  \U_4[\tilde \K]
\ea
is a scalar. 
The replacement hence ensures that the action now formulated as a functional of the metric and the four \stu fields, $S=S[g\mn, \phi^a]$ is explicitly diffeomorphism invariant. Breaking of the symmetry is encoded in the choice of gauge or non-trivial  configuration for the \stu fields; for instance, one can use the explicit gauge freedom to set  $\phi^a = x^a$ which corresponds to  unitary gauge. 
Clearly in unitary gauge $ \partial_\mu \phi^a \partial_\nu \phi^b \eta_{ab}=\eta\mn$ and we recover the previous formulation of massive gravity. In any other gauge, $\tilde \eta$ is still a representation of the Minkowski metric, and we are hence still dealing with a theory of massive gravity on a fixed Minkowski metric. Naturally, one could further decide to consider a different theory, whereby the reference metric is physically different, resulting in a theory which is no longer manifestly Lorentz-invariant in the vacuum (in which the notions of mass and spin may also become more subtle). 

\subsection{Constraints}
Formulated in terms of the \stu fields, massive gravity defined as
\ba
\label{eq:dRGT}
S_{\rm dRGT}[g\mn,\phi^a]&=&\frac{\mpl^2}{2}\int \d^4 x \sqrt{-g}\(R-m^2 \(\tilde \U_2[\tilde \K]+\alpha_3\, \tilde \U_3[\tilde \K]+\alpha_4\, \tilde \U_4[\tilde \K]\)\)\\
&+&\int \d^4x \sqrt{-g} \L_{\rm matter}[g\mn, \psi]\,\nn
\ea
is now manifestly diffeomorphism invariant. So we recover the standard two degrees of freedom in $g\mn$, the standard number of degrees of freedom $\psi_i$ in the matter sector, and now in principle up to four additional degrees of freedom in the \stu fields $\phi^a$. In principle, had we included any arbitrary potential, the gravitational sector of the theory would involve up to six degrees of freedom, namely two degrees of freedom in $g\mn$ and four additional ones in the \stu fields.
The key in ensuring that the dRGT propagates the correct five degrees of freedom (in addition to those encoded in the matter sector) is to ascertain that of the four \stu fields, only three are in fact dynamical. So to make progress, we should consider the dynamics of the  four scalar \stu which are governed by the set of equation of motion, schematically of the form
\begin{equation}
\mathcal{E}_a \equiv \frac{\delta}{\delta \phi^a} \L_{\rm dRGT}[g\mn,\phi^a] = 0 .
\end{equation}
The presence or absence of additional propagating degrees of freedom is encoded in the structure of these equations, and in particular in their dependence on time derivatives of the \stu fields. As shown in \cite{deRham:2011rn}, a useful diagnostic is the Hessian matrix associated with the time derivatives,
\begin{equation}
\mathcal{A}_{ab}
= \frac{\partial \mathcal{E}_a}{\partial \dot{\phi}^b}
= \frac{\delta^2 \mathcal{L}}{\delta \dot{\phi}^a \delta \dot{\phi}^b},
\end{equation}
which determines whether the equations of motion can be solved for the accelerations $\ddot{\phi}^a$.

What makes dRGT massive gravity special (as compared to any other arbitrary mass term), is the fact that this Hessian matrix is degenerate,
\begin{equation}
\det \mathcal{A}_{ab} = 0 ,
\end{equation}
indicating the presence of at least one constraint among the equations of motion. This degeneracy implies that not all \stu fields correspond to independent dynamical degrees of freedom. More explicitly, since the matrix $\mathcal{A}_{ab}$ is null, there must exist a nontrivial vector $c^a$ such that the linear combination
\begin{equation}
\mathcal{C} = c^a \mathcal{E}_a = 0 ,
\qquad
\frac{\partial \mathcal{C}}{\partial \dot{\phi}^a} = 0 ,
\end{equation}
does not involve second time derivatives and thus functions as a second-class constraint, fixing one of the \stu fields to be non-dynamically expressed in terms of the remaining three, instead of providing a dynamical equation of motion, as demonstrated in \cite{deRham:2011rn}. The formulation of this constraint was then extended to the full dRGT structure in \cite{Hassan:2011hr}.

The existence of this primary constraint is a direct manifestation of the special structure of the dRGT interaction potential. Together with an associated secondary constraint, it ensures that one would-be propagating mode is eliminated from the spectrum. Consequently, the full nonlinear theory propagates exactly five degrees of freedom, in agreement with the expectations for a massive spin--2 field \cite{Hassan:2011ea}. This non-perturbative constraint analysis provides a decisive confirmation of the internal consistency of ghost-free massive gravity beyond the decoupling limit and underpins its status as a viable effective field theory.

\subsection{Decoupling Limit}
\label{sec:DL}

One of the clearest ways to further explore the phenomenology of dRGT massive gravity is to look at its decoupling limit. In this regime, the different helicity modes of the massive graviton separate cleanly (\ie decouple), making it much easier to track how each one behaves and diagnose the presence of a constraint. Crucially, the limit preserves the key nonlinear effects that ensure the theory remains consistent and leads to realistic phenomenology, while still allowing the dynamics to be handled analytically within an EFT description. It is defined by taking the limit
\begin{equation}
\label{eq:DLimit}
m \to 0,
\qquad
M_{\mathrm{Pl}} \to \infty,
\qquad
\Lambda_3 = (m^2 M_{\mathrm{Pl}})^{1/3} \;\; \text{fixed},
\end{equation}
so that what will be identified as the strong-coupling scale $\Lambda_3$ remains finite. 
It is worth noting that the decoupling limit \eqref{eq:DLimit} is neither a low-energy limit of the theory (as would be the case in the limit $m/\mpl\to 0$) nor is it a high-energy one. It is a genuinely hybrid limit that captures the first relevant interactions. In this sense, the resulting theory is not a typical low-energy EFT and not all standard low-energy power-counting arguments apply straightforwardly. 
As we shall see, the scale $\Lambda_3$ is the scale at which the nonlinear derivative interactions become important and the theory becomes strongly coupled (which is the essence of the Vainshtein mechanism, precisely as identified by Vainshtein in 1972 by taking the massless limit $m\to 0$ of what was at the time an untuned potential and hence a ghostly theory of massive gravity \cite{Vainshtein:1972sx}).

In this limit, the helicity‑2 sector (now captured in $h\mn$ endowed with linearised diffeomorphism) reduces to the familiar dynamics of linearised GR, while the helicity‑1 and helicity‑0 modes continue to exhibit nontrivial self‑interactions \cite{ArkaniHamed:2002sp,deRham:2010ik}.

Within the decoupling limit, it is particularly straightforward to examine the behaviour of the different degrees of freedom by separating the \stu fields into their individual helicity components.
Without loss of generality, we may express the four \stu fields $\phi^a$ in terms of new variables $A_a$ and $\pi$, 
\begin{equation}
\label{eq:phia}
\phi^a = x^a - \frac{1}{m \mpl}\eta^{ab}\(A_b + \frac{1}{m}\partial_b \pi\) \,,
\end{equation}
where the normalisation scales have been introduced for later convenience. At this stage, this can be interpreted simply as a change of variables, although we will see that these fields $A_a$ and $\pi$ carry a deeper physical significance. 
Since each $\phi^a$ is a scalar under diffeomorphisms, the quantity $A_a$ does not transform as a diffeomorphism vector, and $\pi$ does not behave as a scalar. Nonetheless, in the decoupling limit we will see that they can still be identified with the helicity-1 and helicity-0 modes of the massive spin-2 field, and that their nonlinear interactions are crucial for the Vainshtein mechanism.  
In this decoupling regime, where the dynamics of the different helicity modes can be cleanly separated and effectively mapped onto Lorentzian degrees of freedom, this decomposition furnishes a direct correspondence between massive gravity and the EFT describing interacting scalar and vector fields. \cite{deRham:2010ik,deRham:2011rn,Hinterbichler:2012yr}.

As we shall see, a key outcome of the decoupling‑limit analysis is that the helicity‑0 mode is governed by Galileon‑type derivative interactions. Even though these interactions involve higher derivatives at the level of the Lagrangian, the resulting equations of motion remain second order, avoiding any Ostrogradsky instability \cite{Ostrogradsky:1850fid,Woodard:2015zca}. The connection between the Boulware--Deser ghost and typical Ostrogradsky ghost-like instabilities was indeed cleanly identified in \cite{Deffayet:2005ys}.

Crucially, with the structure of the special mass term identified in \eqref{eq:fullU}, 
the absence of higher derivatives, and hence of any Ostrogradsky instability, ensures that no additional ghostlike degrees of freedom are introduced and that the theory remains consistent in this regime.
The appearance of Galileon interactions is directly related to the structure of the dRGT potential. In the decoupling limit, this structure leads to strong non-‐renormalisation properties and arranges the interactions so that the correct number of degrees of freedom is maintained, even in the presence of significant nonlinearities \cite{Luty:2003vm,deRham:2010ik}. The decoupling limit therefore provides a clean setting in which to test the internal consistency of ghost‑free massive gravity and to understand how infrared modifications of gravity can be realised while retaining theoretical control.

\paragraph{Helicity--0 mode in the Decoupling Limit:} To make progress, we start by focusing on the helicity-0 mode, and switch off the helicity-1 mode for now (it can be safely switched back on later). In this case, the ``St\"uckelbergised" reference metric is given by
\ba
\tilde \eta\mn=\eta\mn-2\tilde \Pi\mn+\tilde \Pi^2\mn = \(\eta-\tilde \Pi\)^2\mn
\ea
where the indices are raised and lowered with the Minkowski metric $\eta\mn$ and we use the notation
\ba
\tilde  \Pi\mn = \frac{1}{\Lambda_3^3}\p_\mu \p_\nu \pi\,.
\ea
We therefore directly see that in the limit \eqref{eq:DLimit} where we send $\mpl\to \infty$ keeping the scale $\Lambda_3$ fixed, we have
\ba
 \lim_{\mpl\to \infty} g^{\mu\nu} = \eta^{\mu \nu } \quad 
\Rightarrow \quad  \lim_{\mpl\to \infty} \tilde \K\mupn = \tilde \Pi\mupn\,.
\ea
The dRGT potential is specially engineered so that the pure helicity-0 interactions, which should be of the form $\Pi^n$ and would be higher order in derivatives, actually combine to give a total derivative and are hence free from the Ostrogradsky instability. For instance, we see that
\ba
\U_2 (\tilde \Pi) & \sim &  \tilde \Pi^\mu{}_\mu \tilde \Pi^\nu{}_\nu - \tilde \Pi^\mu{}_\nu \tilde \Pi^\nu{}_\mu \\
&\sim  &  \frac{1}{\Lambda_3^6}\((\Box \pi)^2 - (\p_\mu \p_\nu \pi)^2\) \sim  \frac{1}{\Lambda_3^6} \p_\mu  \(\p^\mu \pi \Box \pi - \p_\alpha \pi \p^\mu \p^\alpha \pi \)\,.
\ea
 The same logic applies to the other terms $\U_n$ included in the potential. 
So far we have focused on the helicity-0 mode and as we have seen, terms that involve only the helicity-0 mode and no mixing with the helicity-1 or 2 have been specifically designed to be trivial (total derivatives). However, the mass term actually includes an infinite number of interactions
\ba
\L_{\rm mass} & = & -\frac{\mpl^2 m^2}{2} \sqrt{-g}\, \U[g,\tilde \eta] \\
{\rm with }&&  
g\mn=\eta\mn+\frac{1}{\mpl} \tilde h\mn \quad {\rm and }\quad 
\tilde \eta\mn=\(\eta -\tilde \Pi\)^2\mn+\mathcal{O}\(\frac{m}{\Lambda_3^3} \p A\)\,,
\ea
which can be symbolically expressed as an infinite nonlinear expansion
\ba
\left.\L_{\rm mass}\right|_{A=0}  =  -\frac{\mpl^2 m^2}{2} \, \(\U[\tilde \Pi]+\sum_{n\ge 1} \frac{\tilde h^n}{\mpl^n}\(\sum_{k=1}^3 c_k\Pi^k \)\)\,,
\ea
still  omitting the helicity-1 mode (which can be consistently switched off), see \cite{Gabadadze:2013ria,Ondo:2013wka} for the full decoupling limit including those modes. 
As indicated above, the first term $\U[\tilde \Pi]$ is a total derivative and can therefore be safely ignored. In the $\mpl\to \infty$ with the $m\to 0$ limit, the leading order interaction is hence linear in $\tilde h\mn$ which turns out to be \cite{deRham:2010ik,deRham:2010kj}
\ba
\left.\L_{\rm mass}\right|_{A=0}  = \frac {\Lambda_3^3}2 \,  \tilde h ^{\mu\nu}\left[X^{(1)}\mn[\tilde \Pi] +\frac{2+3\alpha_3}{4} X^{(2)}\mn[\tilde \Pi]
+\frac{\alpha_3+4\alpha_4}{4} X^{(3)}\mn[\tilde \Pi]\right] \,,
\ea
with 
\ba
X^{(1)}_{\mu\nu}[\tilde \Pi] &=& [\tilde \Pi]\, \eta_{\mu\nu} - \tilde \Pi_{\mu\nu} \,, \\
X^{(2)}_{\mu\nu}[\tilde \Pi] &=& \big( [\tilde \Pi]^2 - [\tilde \Pi^2] \big) \eta_{\mu\nu} - 2 [\tilde \Pi]\, \tilde \Pi_{\mu\nu} + 2 \tilde \Pi^2_{\mu\nu}, \\
X^{(3)}_{\mu\nu}[\tilde \Pi] &=& \big( [\tilde \Pi]^3 - 3 [\tilde \Pi][\tilde \Pi^2] + 2 [\tilde \Pi^3] \big) \eta_{\mu\nu} \\
&-& 3 \big( [\tilde \Pi]^2 - [\tilde \Pi^2] \big) \tilde \Pi_{\mu\nu} + 6 [\tilde \Pi] \tilde \Pi^2_{\mu\nu} - 6 \tilde \Pi^3_{\mu\nu}\,,\nn
\ea
where square brackets represent the trace of a tensor, \eg $[\tilde \Pi]=\tilde \Pi^\alpha{}_\alpha$, $[\tilde \Pi^2]=\tilde \Pi^\alpha{}_\beta \tilde \Pi^\beta{}_\alpha$, etc.  

\paragraph{Symmetries in the Decoupling Limit:} By construction, the decoupling limit enjoys one local and three global symmetries:
\begin{itemize}
\item Global Lorentz invariance.
\item Linearised diffeomorphism invariance for the helicity-2 modes $\tilde h\mn \to \tilde h\mn+\p_{(\mu}\xi_{\nu)}$.
\item A global shift symmetry on both helicity-0 and -1 modes  \cite{Bonifacio:2019hrj} (maintained in the full decoupling limit including the helicity-1 modes).
\item An additional global Galilean symmetry on the helicity-0 mode $\pi \to \pi+v_\mu x^\mu$ (which manifestly leaves $\Pi\mn$ invariant). 
\item Had we kept the helicity-1 mode $A^a$, it would also have been manifest that the theory exhibits a local $U(1)$ gauge invariance,  $A_\mu\to A_\mu+\p_\mu \chi$. Away from the decoupling limit, the helicity-0 mode should absorb this shift via $\pi \to \pi-m \chi$, but in the decoupling limit $m\to 0$, the symmetry for the helicity-0 and -1 mode separates.  
\end{itemize}
While the linearised diff is only manifest after appropriate integrations by parts, the global Galilean symmetry is directly realised at the level of the Lagrangian by construction. 

\paragraph{Diagonalisation:}
In this formulation, the helicity-0 mode only acquires its kinetic term via mixing with the helicity-2 mode. To properly identify the independent dynamical degrees of freedom and how they couple with external matter fields, it is useful to perform a nonlinear shift and write
\ba
\label{eq:htildeh}
\tilde h\mn=h\mn- \pi \eta\mn-\frac{2+3\alpha_3}{2\Lambda_3^2}\p_\mu\pi \p_\nu \pi\,.
\ea
Note that this shift no longer explicitly preserves the Galilean invariance. So, while the Lagrangian  of massive gravity is explicitly invariant under a Galilean transformation, this is no longer explicit at the level of the Lagrangian when working in terms of the shifted variables (and the invariance  will then only be manifest at the level of the action).

Including the change of variable \eqref{eq:htildeh} in the full dRGT action \eqref{eq:dRGT} and taking the decoupling limit, we obtain the explicit Galileon interactions for the helicity-0 mode
\ba
\label{eq:DL_Gal}
\left.\L^{\rm dec}_{\rm dRGT}\right|_{A=0}&=&-\frac 14 h^{\mu\nu}\hat{\mathcal{E}}^{\alpha\beta}\mn h\ab-\frac 34 (\p \pi)^2+\sum_{n=3}^5
\frac{c_n}{\Lambda_3^{3(n-2)}}\L^{(n)}_{\rm Gal}[\pi]\\
&-&\frac{2(\alpha_3+4\alpha_4)}{\Lambda_3^6}h^{\mu\nu}X\mn^{(3)}+\frac 12 h\mn \tilde T^{\mu\nu}+\frac12\pi \tilde T
-\frac{2+3\alpha_3}{2^4\Lambda_3^3}\p_\mu\pi \p_\nu \pi \tilde T^{\mu\nu}\,,\nn
\ea
where $\tilde T\mn=T\mn/\mpl$ is the (properly normalised) stress-energy tensor associated with the dynamical matter fields $\psi_i$, and the Galileon Lagrangians are given by \cite{Nicolis:2008in}
\ba
\label{eq:Gal2}
\L_{\rm Gal}^{(2)}&=&-\frac{1}{2}(\p \pi)^2\\
\label{eq:Gal3}
\L_{\rm Gal}^{(3)}&=&\frac{1}{2} (\p \pi)^2\, [\Pi]\\
\label{eq:Gal4}
\L_{\rm Gal}^{(4)}&=&\frac{1}{2}\, (\p \pi)^2\, \( [\Pi]^2-[\Pi^2]\)\label{S4nr}\\
\label{eq:Gal5}
\L_{\rm Gal}^{(5)}&=&\frac{1}{2}
(\p \pi)^2 \( [\Pi]^3-3 [\Pi] [\Pi^2]+2 [\Pi^3]\)\,,
\label{S5nr}
\ea
and we have introduced the constant
$c_3=-\frac 34 (2 +3 \alpha_3),
c_4=-\frac 1{8} (4 +9\alpha_3^2+16(\alpha_3+ \alpha_4)),$ and 
$c_5=-\frac 5{16}(2+3 \alpha_3)(\alpha_3+4\alpha_4)$.

\paragraph{Non-renormalisation Theorem:}
Interestingly, the Galileon symmetry is now no longer manifest at the level of the `Galileon' Lagrangians \eqref{eq:Gal2}--\eqref{eq:Gal5}, and only becomes manifest at the level of the equations of motion or effective action. This realisation has important implications on the quantum stability of the resulting action: loops - or quantum corrections -  necessarily generate operators that are \emph{manifestly} Galileon invariant, and hence cannot themselves take a pure `Galileon' form as in \eqref{eq:Gal2}--\eqref{eq:Gal5}. This remarkable feature means that the decoupling limit of massive gravity is in fact protected from quantum corrections, and those only affect higher order operators that enter beyond the decoupling limit \cite{Luty:2003vm,Nicolis:2004qq}. This implies that the tunings that enter the dRGT action are technically natural even though they are not protected by a local symmetry, in the sense that while they get renormalised, they do so by an amount that is technically natural, \ie suppressed by (positive) powers of $m/\mpl$. In the context of dRGT massive gravity, this was explicitly shown in \cite{deRham:2012ew,deRham:2013qqa}.

\subsection{Degrees of Freedom in the Massless Limit of Massive Gravity}

Another essential feature of the theory is the presence of ghost-free higher derivative (kinetic) interactions expressed in \eqref{eq:DL_Gal}. Being able to rely on these interactions is essential for the viability of the Vainshtein mechanism, which is responsible for screening the helicity-0 mode and essential for the phenomenological viability of the theory. Indeed, at the linear level, the helicity-0 mode couples conformally to standard matter field and is hence  responsible for a violation of the equivalence principle. 

\paragraph{vDVZ discontinuity of the linear theory:} Focusing on linear theory, we see in \eqref{eq:DL_Gal} that the helicity-0 mode couples to matter through the coupling $\pi \tilde T$ even in the massless limit $m\to 0$. This feature means that (as far as linear theory is concerned),  when probed via external sources,  the massless limit of massive gravity  $m\to 0$ differs from the massless case $m\equiv 0$ for which that coupling is simply absent. This observation is at the origin of the van Dam Veltman and Zakharov discontinuity \cite{vanDam:1970vg,Zakharov:1970cc}. 

However, as pointed out soon after by Vainshtein, this discontinuity is an artificial feature of the linear theory that is not valid in the small mass limit $m\to 0$ \cite{Vainshtein:1972sx}. Indeed, as is already explicitly shown in \eqref{eq:DL_Gal}, in the small mass limit $\Lambda_3\to 0$ and the nonlinear interactions become increasingly important reaching a strongly coupled regime. In fact, (controlled) strong coupling is essential to the phenomenological viability of the theory \cite{deRham:2014wfa,deRham:2017xox} unless one relies on different mechanisms as proposed in \cite{Gabadadze:2006jm,Gabadadze:2019lld}. Indeed, as shown in \cite{Gabadadze:2019lld} embedding four-dimensional massive gravity into a five-dimensional AdS, enables to rely on the benefits of localized gravity \cite{Bajc:1999mh,Domokos:2015xka,Gabadadze:2021dnk} and massive gravity on AdS \cite{Porrati:2000cp,Porrati:2001db,Porrati:2003sa,Bonifacio:2018zex,deRham:2018axr,Garcia-Saenz:2025htt} to either increase the strong coupling scale \cite{Gabadadze:2017jom} or even resolve both the discontinuity and challenges associated with strong coupling. Such phenomena were also shown to be present about other non-trivial vacua \cite{deRham:2016plk,deRham:2015ijs},  with effects leading to a  higher $\Lambda_2$ strong coupling scale and resolving the usual vDVZ discontinuity.

General lessons from trace-anomaly EFTs also suggest that low strong-coupling scales need not signal a breakdown of the theory. In some cases this indicates the need to include additional degrees of freedom that become important before perturbative unitarity is lost. For instance in Refs.~\cite{Gabadadze:2023quw,Gabadadze:2025uxk}, the local diffeomorphism invariant description of the trace anomaly introduces an additional scalar (the anomalyon/radion), whose interactions soften the pathological growth of amplitudes and provide a weakly-coupled completion up to higher scales, either through spontaneously broken conformal symmetry or its holographic realization \cite{Gabadadze:2023tgi}.

\paragraph{Vainshtein Screening:}

The relevance of strong coupling  in the small mass limit of gravity 
was identified by Vainshtein in 1972, but only in 2000 was the first realisation of this mechanism fully explored within the context of the Dvali-Gabadadze-Porrati (DGP) model of soft massive gravity (resonance) \cite{Dvali:2000hr}. A key aspect of the Vainshtein screening mechanism is the existence of a regime in which the helicity-0 mode effectively becomes Galileon-like, exhibiting higher-derivative interactions while remaining free of ghost instabilities. Due to these nonlinearities, the helicity-0 mode can form a condensate and, in practice, become completely decoupled from the remaining gravitational and matter sectors. This is made explicit in the decoupling limit of massive gravity: the helicity-0 mode not only emerges as an independent degree of freedom, but also ceases to interact with any gravitational or matter field in the small-mass limit $m \to 0$. The ghost-free structure of the interactions in this regime is therefore crucial, as it allows one to exploit these nonlinear operators consistently and reorganise the usual EFT power-counting scheme. 

To see this point, the simplest is to focus on the simple cubic Galileon (which can  mimic the leading interactions of the helicity-0 mode of dRGT massive gravity or that from more general  models of soft massive gravity from extra dimensions),
\ba
\label{eq:cubicGal}
\L=-\frac12 (\p \pi)^2 +\frac{1}{\Lambda_3^3}(\p \pi)^2 \Box \pi + \frac{1}{\mpl}\pi T\,.
\ea
The precise implementation of the Vainshtein mechanism in generic (non-symmetric) configurations is still under investigation, although numerical progress has proven to be successful \cite{Gerhardinger:2024rza,deRham:2024xxb}. For here we shall simply focus on the static and spherically symmetric configuration $\pi=\phi(r)$ assuming  a non-relativistic spherically symmetric body localised at $r=0$, so that  $T=-M \delta^{(3)}(r)$. The magnitude of the force $\phi'(r)$ mediated by the helicity-0 mode $\phi$ then enjoys the following behaviour,
\ba
\phi'(r)\sim \begin{cases}
\frac{M}{\mpl}\frac{1}{r^2} &\mbox{for } r \gg r_* \\
\frac{M}{\mpl}\frac{1}{r_*^{3/2}r^{1/2}} & \mbox{for } r \ll r_*
 \end{cases} \,.
\ea
where we have defined the Vainshtein radius associated with the source $M$ as $r_*^3=M/(\mpl\Lambda_3^3)$. 

In particular, this implies that at large distances compared with this Vainshtein radius, $r \gg r_*$, the linear regime is good and the helicity-0 mode propagates an additional Newton-type inverse-square-law force. In this regime, corrections to GR are of order one due to the presence of an unscreened, fully active additional polarisation for gravity, and hence a fifth force. 

However, within this so-called Vainshtein radius, $r \lesssim r_*$, nonlinearities become important, and this is precisely how the Vainshtein mechanism takes effect, suppressing the effect of the helicity-0 mode. Indeed, we can see that well within the Vainshtein radius, the force mediated by this mode is suppressed compared to the standard gravitational force,
$\phi'/F_{\rm Newton} \sim (r/r_*)^{3/2}$ and vanishes as $r\to 0$ or $r_* \to \infty$, so the helicity-0 mode no longer propagates any force in that limit, meaning that it becomes effectively decoupled. This is precisely what occurs in the massless limit of massive gravity, since in this case $r_*\to \infty$, irrespective of the mass $M$. Even though technically speaking the helicity-0 mode is still present in the massless limit of massive gravity, `observing' it or `interacting' with it would require a non-zero (even if infinitesimally small) probe $M$ which would plunge the whole Universe within its own Vainshtein radius and fully decouple the effect of this helicity-0 mode.   

Another aspect that is relevant to highlight in this limit is that even though the Vainshtein mechanism relies on strongly coupled interactions, this mechanism raises the strong coupling scale of perturbations. Indeed, considering fluctuations on top of this static and spherically symmetric condensate, $\pi= \phi(r)+\delta \pi(x^\mu)$, the background condensate redresses the kinetic term of the fluctuations. Symbolically, 
\ba
\L_{\delta \pi}\sim \(1+\frac{\phi'}{r\Lambda_3^3}\)(\p \delta \pi)^2+\frac{1}{\Lambda_3^3}(\p \delta \pi)^2 \Box \delta \pi+\frac{1}{\mpl}\delta \pi \delta T\,.
\ea
Canonically normalising the field, symbolically $\delta\widehat{ \pi}=\delta \pi /\sqrt{Z}$, with $Z\sim \(1+\frac{\phi'}{r\Lambda_3^3}\)$, we then see that the dynamical degree of freedom  on this condensate behaves as 
\ba
\L_{\delta\widehat{ \pi}}\sim\(\p \delta\widehat{ \pi}\)^2+\frac{1}{\tilde \Lambda_3^3}(\p \delta\widehat{ \pi})^2 \Box \delta\widehat{ \pi}+\frac{1}{\sqrt{Z}\mpl}\delta\widehat{ \pi} \delta T\,.
\ea
with $\tilde \Lambda_3=\sqrt{Z}\Lambda_3$ and $Z\gg 1$ in the Vainshtein radius. This implies that the dynamical field $\delta\widehat{ \pi}$ becomes weakly coupled to matter to the point that it decouples in the limit $m\to 0$ and the new strong coupling scale $\tilde \Lambda_3$ has been raised, and in terms of mass $m$, it scales as $\tilde \Lambda_3\sim m^{1/6}$. 

A particular feature of Galileons with the Vainshtein mechanism and of massive gravity is the presence of superluminalities for perturbations about a non-trivial profile  \cite{Luty:2003vm,Adams:2006sv,Nicolis:2008in,Gruzinov:2011sq,deRham:2011pt,Deser:2012qx}. The connection between these superluminalities and causality  violation has proven challenging to confirm, see Ref.~\cite{Burrage:2011cr} for more detail as well as the discussion on front velocity in \cite{deRham:2014zqa}, nevertheless, these features are certainly non-standard and may indicate the need to embed these classes of theories onto non-standard high-energy completion \cite{Dvali:2010jz,Aydemir:2012nz,deRham:2014wnv,Keltner:2015xda,deRham:2018bgz}. When the Galileon is treated as an independent scalar field, this expectation is indeed verified in \cite{Tolley:2020gtv}. However, applying analogous arguments in setups where the Galileon is tied to the spin-2 pole necessitates a fully gravitational analysis, and thus the overall implications for massive gravity remain, so far, not entirely settled.

\vspace{0.3cm}
\noindent \textbf{\color{color1}Bottom Line:} Operators that enter dRGT massive gravity are uniquely 
determined by the requirement that the theory propagates five ghost-free  degrees of freedom non-perturbatively (in four dimensions). In the decoupling limit, the helicity-0 mode 
acquires Galileon-like type of  interactions that are protected from quantum corrections via a non-renormalisation theorem and 
enable a Vainshtein screening mechanism which essential for the phenomenological viability of the theory.

\section{Observational Bounds on the Graviton Mass}

As discussed in the previous section, the Vainshtein mechanism is essential for maintaining a certain degree of phenomenological viability in massive gravity. In addition, the dRGT theory contains only two free parameters beyond the overall mass scale. For the theory to remain phenomenologically viable, these coefficients must therefore be selected so that a Vainshtein mechanism can operate (though not necessarily in exactly the same way as in the static, spherically symmetric solutions discussed above). Moreover, the mass cannot be too large, otherwise the Vainshtein mechanism would not work efficiently enough. By now, there exist a multitude of observational constraints on the graviton mass and the two free coefficients. 
Since massive gravity is a modification of gravity in the infrared, naturally the strongest bounds arise from observations sensitive to long-range gravitational physics or to the propagation of gravitational waves over long astrophysical and cosmological distances \cite{deRham:2016nuf}.

Phenomenological constraints on the graviton mass $m$ (or equivalently on its Compton wavelength $\lambda_m$) can be separated into three complementary categories depending on the actual modification probed (eV stands for electronvolt and ly for lightyear).
\begin{itemize}
\item {\bf Yukawa potential:} The original motivation behind massive gravity is to weaken its behaviour in the infrared so as to directly tackle the cosmological constant problem. Nominally, the effect one is after is an effective modification of the gravitational potential in the infrared. In the non-relativistic limit, and if one were able to work at the linear level, the Newtonian/Coulomb potential would be modified to include a Yukawa exponential suppression.  Based on this assumption, the constraint on the graviton mass can range from $m\lesssim 10^{-23}$eV, ($\lambda_m\gtrsim \mathcal{O}(1)$ly $\sim10^{13}$km) from the precession of Mercury \cite{Williams:2004qba} all the way down to $10^{-32}\mathrm{eV}$, ($\lambda_m\gtrsim 10^9$ly$\sim10^{22}$km) from the convergence scale of the CMB dipole 
\cite{Loeb:2024wkv}. Although constraints on the graviton mass from a Yukawa-like potential are typically the strongest, they rely on the very strong linear-regime assumption, which is generically not valid given the importance of nonlinear effects in massive gravity. 
\item {\bf Fifth force:} As discussed previously, one of the most drastic features of massive gravity is the presence of additional polarisations and in particular the helicity-0 mode, which propagates a `fifth' force. The Vainshtein mechanism typically ensures the screening of this mode for sufficiently small values of the mass (or related scale). However, fifth-force constraints and tests of the equivalence principle are extremely precise.
Bounds on the graviton mass coming from fifth-force constraints are typically quite model dependent as they depend strongly on the precise details of the nonlinear Vainshtein screening mechanism. In the case of DGP, the Lunar Laser Ranging experiment places an extremely tight constraint on the scale of the graviton resonance $m\lesssim 10^{-32}$eV, ($\lambda_m\gtrsim 10^9$ly$\sim10^{22}$km) \cite{Merkowitz:2010kka,Dvali:2002vf}, while binary pulsar systems put constraints on that same scale to be  $m \lesssim 10^{-27}~\mathrm{eV}$, ($\lambda_m\gtrsim 10^4$ly$\sim10^{17}$km)    \cite{Finn:2001qi}. Constraints from fifth-force experiments are typically less competitive for sharper resonances, like Cascading Gravity or for hard massive gravity as in dRGT since the Vainshtein mechanism relies on higher-order operators. If the interaction of the helicity-0 mode with the other Standard Model fields leads to an additional violation of the equivalence principle, next-generation atomic clocks may be able to set competitive bounds on the graviton mass \cite{Elder:2025tue}.
\item {\bf Dispersion Relation:} Finally, the cleanest (least model-dependent) bound on the graviton mass comes from the effect that the mass has on the dispersion relation of standard (helicity-2) gravitational waves. The leading order corrections are of the form
\begin{equation}
E^2 = {\mathbf{k}}^2 + m^2 \,, \quad {\rm or} \quad v_g^2(E)=1-\frac{m^2}{E^2} \, .
\end{equation}
 This conclusion holds for broad classes of massive gravity theories, including resonances, where an equivalent assertion can be formulated directly in terms of the spectral representation. This kind of effect is hence generic to any theory of gravity which genuinely affects the dynamics of the actual gravitational degrees of freedom in the infrared and can be probed directly at the linear level. Because it relies only on the dynamics of the helicity-2 mode, these tests are insensitive to the screening mechanism and other nonlinear subtleties of the theory. Another advantage of these kinds of modifications is that their effects accumulate over large distances and leave a signature on gravitational wave signals, either as phase shifts, distortions of the waveform, or changes in arrival times in multi-messenger or multi-frequency observations. Observations by the LIGO--Virgo--KAGRA collaboration, including binary black hole and binary neutron star mergers, place robust upper bounds on the graviton mass by constraining deviations from luminal propagation \cite{Abbott:2016blz,Abbott:2017xzu}. Current data imply
\ba
m &\lesssim& 10^{-22}~\mathrm{eV}\,,\\
\lambda_m &\gtrsim& 10^{-1}~\mathrm{ly} \sim 10^{-2}{\rm pc}\sim 10^{12}{\rm km},
\ea
corresponding to a Compton wavelength exceeding typical astrophysical scales. Although these bounds are not as competitive as other phenomenological constraints on massive gravity, they are robust in the sense that they apply to a broad class of models and rely on limited assumptions. 
\end{itemize}
Note that all these bounds are still consistent with the graviton mass being of the order of the Hubble parameter today, which is the relevant regime of interest for cosmology and for tackling the cosmological constant problem. 

\vspace{0.3cm}
\noindent\textbf{\color{color1}Bottom Line:} Current observational constraints on the graviton mass can be classed into three main categories: 1. bounds arising from Yukawa-like modifications of the gravitational potential, 2. constraints from fifth-force searches, and 3. constraints from the dispersion of gravitational waves. Among these, gravitational-wave observations provide the cleanest and most robust tests, currently leading to bounds that remain compatible with the parameter range relevant for addressing the cosmological constant problem.

On the other hand, the first two classes of constraints is  more subtle due to the theory's kinetic nonlinear structure. In particular, the presence of a constraint in dRGT massive gravity forbids solutions that are simultaneously static and spherically symmetric  or fully homogeneous and isotropic. As a result, phenomenologically viable configurations typically require either a mild time dependence (on Cosmic scale) or a departure from exact homogeneity and isotropy on cosmological scales. While such dependencies are fully phenomenologically viable in principle it challenges the use of standard tools. 
Getting more definitive observational signatures, therefore requires a deeper understanding of the relevant nonlinear solutions and their full phenomenology.

\section{Theoretical Consistency}

A central property of massive gravity is the existence of a non-linear constraint that prevents the existence of local solutions with high degrees of symmetry, including perfectly homogeneous and isotropic solutions relevant to cosmology \cite{D'Amico:2012zv}, as well as static, spherically symmetric configurations important for black hole physics.\cite{Berezhiani:2008nr,Berezhiani:2011mt,Deffayet:2011rh}. A review of exact solutions in massive gravity relevant for cosmology and black hole physics was provided in \cite{Volkov:2013roa} (see also Ref.~\cite{Gervalle:2020mfr,Hogas:2022owf} for more recent work). For cosmology, mildly breaking homogeneity on distance scales of order the graviton Compton wavelength (as encoded by the \stu fields) may provide more realistic solutions, but such configurations have remained challenging to construct analytically. An analogous situation arises for black hole solutions, where a mild time dependence (on time scales set by the graviton mass) was shown to be generic \cite{Rosen:2017dvn,Rosen:2018lki}, yet exact analytic solutions that smoothly connect the black hole horizon to asymptotic infinity are likewise hard to obtain. The viability of the theory clearly hinges on the phenomenology of these solutions, and analytic progress has remained a major obstacle. This, in turn, motivates the development of more robust numerical methods for studying such configurations, as discussed below, where some parallel progress has been achieved in recent years.

\subsection{A Well-Posed Dynamical Formulation}

A long-standing difficulty in the nonlinear study of massive gravity, as with many EFTs of gravity, has been its formulation as a manifestly well-posed initial value problem. Although ghost-free massive gravity is known to propagate the correct number of degrees of freedom, non-perturbatively, \ie around any background, the explicit construction of a system of differential equations suitable for time evolution via numerical simulations has remained subtle. The difficulty does not stem from the presence of additional modes, but rather from the highly nonlinear constraint structure combined with the explicit breaking of spacetime diffeomorphism invariance, which obscures both constraint propagation (since constraints are no longer associated with symmetries) and the straightforward identification of a suitable principal symbol.

Nevertheless, it is worth emphasising that the subtleties related to well‑posedness and the numerical evolution of time are not indications of any fundamental theoretical inconsistency. Rather, these features are standard in generic EFTs, including the Navier–Stokes formulation of fluid dynamics and the low‑energy Euler–Heisenberg EFT. In these well‑understood examples, such issues simply reflect the fact that the effective description is being pushed beyond its regime of validity, signalling the need to incorporate additional ingredients from the UV completion once the dynamics probe scales near or above that of strong coupling. In the case of Navier–Stokes, the underlying molecular or granular nature of the fluid inevitably comes into play on sufficiently small scales, providing a physical regulator that naturally resolves the apparent breakdown of the low-energy dynamics. 

Just as well‑posedness or numerical evolution in Navier–Stokes or Euler–Heisenberg do not represent any fundamental inconsistency, but rather an expected and benign limitation of the EFTs, one can expect the same to happen in many other low-energy EFTs and indeed recent progress has demonstrated that these obstacles are also easily surmountable in massive gravity and related theories.
In particular, by reformulating ghost-free massive gravity in terms of vielbein-like variables \cite{Hinterbichler:2012cn} one obtains a framework in which the constraint structure becomes more transparent, and the evolution equations admit a well-defined and robust initial value formulation
\cite{deRham:2023ngf,Kozuszek:2024vyb,Albertini:2024kmf} (see also \cite{Gerhardinger:2024rza} for a UV-complete numerical formulation of quartic Galileons relevant for massive gravity). 
Typically, a vielbein $E^\alpha{}_\mu$ (or tetrad) is a field that relates the spacetime coordinate basis (and dynamical metric $g\mn$) to a locally inertial frame (with fixed Minkowski metric $\eta_{\alpha \beta}$)  $g\mn=E^\alpha{}_\mu E^\beta{}_\nu \eta_{\alpha \beta}$. Effectively, this is taking the `square root' of the metric which is why such a variable is so convenient in the context of massive gravity. Here we shall promote that definition to an arbitrary reference metric $f_{\alpha \beta}$.
This construction provides the first fully nonlinear demonstration that massive gravity can be formulated as a strongly well-posed dynamical system, suitable for controlled time evolution.

For any reference metric  $f\ab$, the dynamical spacetime metric can be written in terms of the vielbein $E_{\alpha \mu}$ as
\begin{equation}
g_{\mu\nu} = (f^{-1})^{\alpha\beta} E_{\alpha\mu} E_{\beta\nu}\,.
\end{equation}
This formulation works for all parameters of dRGT massive gravity, but focusing on a specific choice for simplicity, and expressing the action in the vielbein formalism, we have
\begin{equation}
S
=
\frac{\mpl^2}{2}
\int \d^4x \sqrt{-g}
\left[
R[g]
-
m^2\([E^2]-[E]^2+6 [E]-12\)
\right]
+
S_{\mathrm{matter}}[g,\psi_i],
\end{equation}
where, as before, square brackets represent the trace of the enclosed tensor. 
As we shall see, this formulation is a convenient one as it makes the constraint algebra manifest and stable under time evolution, while clearly isolating the non-dynamical variables responsible for enforcing constraints \cite{Kozuszek:2024vyb,Albertini:2024kmf}.

\paragraph{Phase-space Variables and the Scalar Constraint:}

To proceed we may introduce non-standard momenta for the spatial components of the vielbein,
\begin{equation}
P_i \equiv \partial_t E_{it},
\qquad
P_{ij} \equiv \partial_t E_{ij}\, . 
\end{equation}
The magic of ghost-free massive gravity theories, contained in the special mass terms, is that the temporal component $E_{tt}$ does not acquire a conjugate momentum. This feature is central to the consistency of the formulation and leads to the scalar constraint responsible for eliminating the Boulware--Deser ghost.

What is special about this formalism is that the constraint remains purely algebraic in $E_{tt}$ and does not involve time derivatives.
As a result, $E_{tt}$ can be solved algebraically on each time slice rather than being determined by an evolution equation. Constraint preservation therefore does not generate higher time derivatives or secondary dynamical equations. Importantly, this result holds fully nonlinearly and does not rely on the decoupling limit or perturbative expansions \cite{deRham:2023ngf,Kozuszek:2024vyb} and hence provides a relatively simple way to demonstrate the correct number of propagating degrees of freedom.

\paragraph{Structure of the Evolution Equations:}

After solving the algebraic constraint and isolating the true dynamical variables, the equations of motion can be cast into a first-order-in-time system of the schematic form
\begin{equation}
\partial_t \tilde{P}_{ij} = S_{ij},
\qquad
\partial_t \tilde{E}_{ij} = u_{ij},
\qquad
\partial_t E_i = v_i,
\qquad
\partial_t \tilde{E} = W,
\end{equation}
where tildes denote combinations in which the constraints have been explicitly solved.
The leading spatial derivative terms appearing in $S_{ij}$ take the form
\begin{equation}
S_{ij}
=
J_{ij}{}^{klmn}\,\partial_k\partial_l \tilde{E}_{mn}
+
J_{ij}{}^{klm}\,\partial_k\partial_l \tilde{E}_m
+
J_{ij}{}^{kl}\,\partial_k\partial_l \tilde{E}
+
\cdots ,
\end{equation}
with analogous expressions for $u_{ij}$, $v_i$, and $W$. The coefficient tensors $J$ depend only algebraically on the fields and not on their derivatives, and this allows the principal symbol of the system to be unambiguously identified \cite{deRham:2023ngf,Albertini:2024kmf}.

\paragraph{Diffusion Terms and Well-Posedness:}

In the case of minimal massive gravity, it has recently been shown to be strongly hyperbolic for generic metrics sufficiently
close to Minkowski, so that its Cauchy evolution problem is well-posed \cite{Kozuszek:2024vyb}. There is an expectation that this holds more generally, although this has not been fully confirmed at present. However, remembering that massive gravity should be understood as an effective field theory, to guarantee strong well-posedness in the sense of Hadamard, it is sufficient to introduce diffusion terms,
\ba
\begin{array}{rcl}
\partial_t \tilde{P}_{ij}
&=&
S_{ij}
+
\ell^2 \delta^{mn}\partial_m\partial_n \tilde{P}_{ij},
\\
\partial_t \tilde{E}_{ij}
&=&
u_{ij}
+
\ell^2 \delta^{mn}\partial_m\partial_n \tilde{E}_{ij},
\\
\partial_t E_i
&=&
v_i
+
\ell^2 \delta^{mn}\partial_m\partial_n E_i,
\\
\partial_t \tilde{E}
&=&
W
+
\ell^2 \delta^{mn}\partial_m\partial_n \tilde{E}.
\end{array}
\ea
These terms act as a numerical regulator, improving ultraviolet conditioning without modifying the physical content of the theory. Physical solutions remain unaffected in the regime
\begin{equation}
T \ll \ell^{-2},
\qquad
L \gg \ell,
\end{equation}
where $T$ and $L$ denote the characteristic time scale and length scale of the configuration.

Importantly, as emphasised in \cite{deRham:2023ngf,Kozuszek:2024vyb}, the existence and uniqueness of solutions are insensitive to the precise value of $\ell^2$, demonstrating that well-posedness is not an artefact of the regulator, but an intrinsic property of the formulation. We can think of the energy scale at which the regulator comes in as the cutoff of the EFT. If we can successfully show that all low energy physics is insensitive to the regulator for a sufficiently large hierarchy $L/\ell$ then we can declare the predictions of the simulation meaningful and the evolution well-posed. Closely related approaches for Galileon theories that arise in the decoupling limit are given in \cite{Gerhardinger:2024rza,deRham:2024xxb}.

\paragraph{High-Frequency Analysis:}

Well-posedness is established by examining the principal symbol. Linearising around a background solution and inserting plane-wave perturbations,
\begin{equation}
\delta X \sim e^{-\omega t + i k_m x^m},
\end{equation}
one finds that at high spatial frequencies the evolution equations reduce to
\begin{equation}
\partial_t
\begin{pmatrix}
\delta \tilde P_{ij}\\
\delta \tilde E_{ij}\\
\delta E_i\\
\delta \tilde E
\end{pmatrix}
=
\left(
\begin{array}{cccc}
\ell^{2}\,\delta_i^{m}\delta_j^{n}\delta^{kl}
& J_{ij}{}^{klmn}
& J_{ij}{}^{klm}
& J_{ij}{}^{kl}
\\[3pt]
0
& \ell^{2}\,\delta_i^{m}\delta_j^{n}\delta^{kl}
& 0
& 0
\\[3pt]
0
& 0
& \ell^{2}\,\delta_i^{m}\delta^{kl}
& 0
\\[3pt]
0
& 0
& 0
& \ell^{2}\,\delta^{kl}
\end{array}
\right)
\,\partial_k\partial_l
\begin{pmatrix}
\delta \tilde P_{mn}\\
\delta \tilde E_{mn}\\
\delta E_m\\
\delta \tilde E
\end{pmatrix}
+\cdots \,.
\label{eq:paris85_matrix}
\end{equation}
This yields the dispersion relation
\begin{equation}
\omega = \ell^2 k^2 ,
\end{equation}
demonstrating parabolic behaviour at high frequencies and ensuring strong well-posedness, with existence, uniqueness, and continuous dependence on initial data. Crucially, this behaviour is independent of the detailed ultraviolet numerical completion: the regulator affects only numerical conditioning, not the mathematical or physical consistency of the evolution \cite{deRham:2023ngf,Albertini:2024kmf}.

\paragraph{Interpretation:}

Although we have focused on a special choice of parameters for the sake of concreteness, the same logic follows for all parameters of dRGT massive gravity \cite{Kozuszek:2024vyb,Kozuszek:2025lem}. This procedure and the existence of an explicitly well-posed formulation of the theory show that ghost-free massive gravity admits a fully nonlinear dynamical system with an algebraic scalar constraint, where the propagation of the  constraint can be followed in a controlled way. 
Although this result is still relatively recent, it now  opens the door to the study of fully nonlinear simulations, including in strong-field regimes which can be applied to find black hole and  cosmological solutions in massive gravity, something that has been a longstanding stumbling block until now \cite{Albertini:2024kmf}. Separately, together with the related approaches of \cite{Gerhardinger:2022bcw,Gerhardinger:2024rza,deRham:2024xxb} for Galileons, they show that theories that exhibit the Vainshtein screening mechanism can be put on a firm numerical footing.

\subsection{A Local UV Completion?}

So far we have established the  stability of Massive Gravity as a low-energy EFT (absence of gradient, ghost, and tachyonic instabilities and well-posed formulation) and have identified which phenomenological tests could provide the best observational signatures. However, even if a theory is fully consistent with existing observations and experiments and represents a stable low-energy EFT,  it may still not admit a well-defined standard and local UV completion. 
General arguments show that there cannot be a Higgs mechanism for gravity with a finite number of degrees of freedom \cite{Kakushadze:2000zn} in Minkowski. In \cite{Bonifacio:2019mgk} it was also shown that massive gravity cannot admit a tree-level UV completion with a finite number of new lower-spin modes but the situation may differ with higher-spin states\cite{Boulanger:2006tg} or on Anti-de Sitter \cite{Porrati:2001db}.
Here by `standard' UV completion, we have in mind a local embedding of the theory at arbitrarily high energies, (not necessarily a QFT described with a well-defined field content), but which is
\begin{enumerate}
\item Poincar\'e invariant,
\item Unitarity, in the sense that the $S$-matrix satisfies $S^\dagger S=1$,
\item Causal, in the sense that the $S$-matrix is an analytic function aside from cuts and poles on the real axis~\cite{bogoliubov1959introduction,Hepp_1964,Bremermann:1958zz,deRham:2017zjm},
\item Local, in the sense that the Froissart bound is satisfied \cite{Froissart:1961ux}, and more generally that the scattering amplitudes are polynomially or exponentially bounded.
\end{enumerate}
Among these assumptions, the last one is not as well established for gravitational EFTs, and most subtleties in applying this logic to gravity are limited by the last point \cite{Alberte:2020jsk}. 
In the context of massive gravity, one would naturally expect it to enjoy a different class of UV completion (\eg see \cite{deRham:2018bgz} and the introduction of \cite{deRham:2018qqo} for a broader discussion on more generic expectations), but for now we shall nevertheless follow through with these assumptions and infer their implications for massive gravity as a standard EFT. 

With these considerations in mind, we can consider the elastic  $2\to 2$ scattering of states in our theory. Note that we do not necessarily need to consider scattering of gravitons, scattering any state allowed in the theory still provides useful insights on the interactions and consistency of the theory. 
Denoting by $\mathcal{A}$ the scattering amplitude and by $s,t,u$ the standard Mandelstam variables;  ($s$ being the centre-of-mass energy squared, $t$ the momentum transfer, and $u=4m^2-s-t$, where $m$ is the mass of the field being considered, an extension of these variables can be used when scattering particles with different masses \cite{Alberte:2019lnd,Alberte:2019xfh,deRham:2025htd}). In terms of the scattering angle $\theta$, the momentum transfer is
$t=-(s-4m^2)(1-\cos \theta)/2$.
A remarkable result is that unitarity (through the optical theorem), together with the other previous assumptions on the UV completion, implies properties of the amplitude directly applicable at low energies \cite{Adams:2006sv}
\ba
\label{eq:Pos1}
\sigma (s) =\frac{{\rm Im}\mathcal{A}(s,0)}{\sqrt{s(s-4m^2)}}>0\,, \qquad \text{in the physical region }\quad  s\ge 4m^2\,.
\ea
\paragraph{Further Bounds:}
This first `positivity bound' \eqref{eq:Pos1} is the first in an infinite set of positivity constraints. Indeed, using the partial wave expansion of the amplitude,
\ba
\mathcal{A}(s,t)=16 \pi \sqrt{\frac{s}{s-4m^2}}\sum_{\ell=0}^\infty (2\ell+1)P_\ell (\cos \theta)a_\ell(s)\,,
\ea
full unitarity can be imposed directly at the level of each coefficient of the partial wave expansion (still in the physical region)
\ba
0\le |a_\ell(s)|^2\le {\rm Im}a_\ell(s)\le 1\qquad{\rm for}\qquad s\ge 4m^2\,.
\ea
Recasting this statement at the level of the amplitude in terms of the scattering amplitude demands positivity of \cite{deRham:2017avq}
\ba
\label{eq:ptn}
\frac{\p^n}{\p t^n}{\rm Im}\mathcal{A}(s,t)>0\,, \quad \forall \ n>0 \quad && {\rm for}\quad s\ge 4m^2\\
\nn {\rm and} && {\rm for}\quad 0\le t <m^2\,.
\ea

\paragraph{Local UV Constraints on Massive Gravity:}
We started this review treating GR as an EFT with cutoff at (or below) the Planck scale. Similarly, we can treat  massive gravity as a low-energy EFT with a cutoff that could in principle be as low as $\Lambda_5=(\mpl m^4)^{1/5}$. In the special case of dRGT massive gravity, the cutoff\footnote{The existence of a Vainshtein mechanism requires the excitation of non-trivial operators at the scale $\Lambda_3$, so from that perspective, it is essential to treat $\Lambda_3$ as the strong coupling scale and not the cutoff for the Vainshtein screening mechanism to work. In this sense, the logic behind the Vainshtein mechanism differs from the standard EFT approach and requires a non-trivial reorganisation of the operators, see \cite{deRham:2014wfa,deRham:2014wnv,deRham:2014fha} for more details. In this section, we solely focus on the standard local UV completion approach where the Vainshtein redressing and re-organisation of the EFT is not accounted for.} can be raised to the scale $\Lambda_3=(\mpl m^2)^{1/3} \gg \Lambda_5$ \cite{deRham:2010kj} (or even as high as $\Lambda_2=(\mpl m)^{1/2}$ on AdS or other non-trivial vacua \cite{deRham:2016plk,deRham:2015ijs,Gabadadze:2017jom,deRham:2018svs}). 

Setting these subtleties aside for now, the standard EFT approach would suggest to start by expressing the low-energy EFT of a massive spin--2 field as an operator expansion, with operators entering at the scale $\Lambda_3$ and those entering at lower-energy scales $\Lambda_5=(m^4 \mpl)^{1/5}$ and $\Lambda_4=(m^3 \mpl)^{1/4}$, where 
\ba
m \ll \Lambda_5 \ll \Lambda_4 \ll \Lambda_3 \ll \Lambda_2 \ll \mpl\,.
\ea
Focusing on the operators that are quartic in the fields, so they enter the $2 \to 2$ scattering amplitude at tree-level, we have symbolically
\ba
&& \L=\L_{\rm \Lambda_3}+\L_{\Lambda_5-{\rm operators}}+\L_{\Lambda_4-{\rm operators}}+\cdots\,,\\
&& \L_{\Lambda_5-{\rm operators}}+\L_{\Lambda_4-{\rm operators}}=\Delta c [h^3]+\Delta d [h^4]
\ea
By considering scatterings between states of appropriate mixed helicities (encoded with the mixing parameters $\alpha_{\pm 1}$) as well as beyond the forward limit positivity bounds $t> 0$,  unitarity, causality, Poincar\'e invariance and locality of the UV completion imply
\cite{deRham:2018qqo}
\ba
\(\alpha_{+1}^2-\alpha_{-1}^2\)\Delta c &>&0\\
{\rm and}\quad
\frac{m^2}{\Lambda_5^{10}}\Delta d\(s-2m^2+\frac t 2\)+\mathcal{O}\(\frac{m^4}{\Lambda_5^{10}}\)&>&0\,.
\ea
These bounds should be satisfied for all helicity choices (\ie for all $\alpha_{j}$ with $\sum_j |\alpha_j|^2=1$) as well as for all momentum transfers $0<t<m^2$.
So as far as the $2 \to 2$ scattering amplitude is concerned,  unitarity, causality, Poincar\'e invariance and locality of the UV completion requires any theory of a massive spin--2 field to have $\Delta c=\Delta d=0$ and hence to have no operators arising at a scale below $\Lambda_3$. 
This is quite a remarkable result in its own right as it indicates that  any massive spin--2 EFT needs to carry the ghost-free ($\Lambda_3$) parameters (at least for what the $2 \to 2$ scattering is concerned) to be even potentially embeddable in a  standard and local UV completion \cite{deRham:2018qqo}.

Positivity of the amplitude arising from the  $\Lambda_3$ operators (corresponding to the ghost-free operators), was first explored in  \cite{Cheung:2016yqr} where an island of positivity was identified in the forward limit and shown to remain positive including higher order bounds  beyond the forward limit  \cite{deRham:2018qqo}. This island contains regions of parameter space that can exhibit the Vainshtein mechanism and be phenomenologically viable. This result is very surprising in its own right as the very logic of massive gravity, relying on a strong coupling redressing and reorganisation of the EFT approach would naively have suggested to prevent the existence of a standard local UV completion \cite{Keltner:2015xda}. Indeed, massive gravity is expected to enjoy a slightly different and perhaps richer kind of UV completion as we shall see in Section~\ref{sec:TTbar}. 

\paragraph{Improved Positivity Bounds:} 
Maintaining a standard EFT approach for now, the previous results indicate that the ghost-free (dRGT) structure of massive gravity should not only be understood as the \emph{maximally soft} infrared realisation of a massive spin--2 theory but in fact as the preferred realisation potentially capable of enjoying a standard local UV completion.

When treated as an EFT with cutoff $\Lambda$ which may differ from the strong coupling scale $\Lambda_3$, \emph{improved positivity bounds} enable us to subtract the contribution to the branch cut from physics below the cutoff (for which the EFT should be under control).  A violation of any of these bounds would indicate that new degrees of freedom, inelastic channels, or non-local effects must enter at or below the corresponding scale. In  \cite{Bellazzini:2017fep}, these considerations were used to show that under a sufficiently weak coupling assumption (\ie assuming sufficiently small $g\ll 1$, where $g$ is the coupling constant that effectively plays the role of $\hbar$ determining the overall higher-loop sensitivity) the improved bounds can rule out the physically relevant regions of parameter space. However, as indicated in \cite{deRham:2017imi}, these improved non-forward positivity bounds can and should, in fact, be more appropriately understood as determining the value of the cutoff and the onset of the strong coupling, indicating, as anticipated, the emergence of pronounced strong-coupling effects in the theory. 

\paragraph{Bounds at negative $t$:} Further studies involving positivity bounds at negative $t$ (beyond the physical region) were explored in \cite{Bellazzini:2023nqj}. The analysis uses dispersion relations and the partial wave expansion to infer a relation on the finite $t<0$ scattering amplitude. Given a helicity scattering amplitude ${\cal A}_{\lambda_1 \lambda_2}(s,t)$ for which the out states have the same helicities as the in states and defining\footnote{Note that this differs from what is included in \cite{Bellazzini:2023nqj} as the authors  neglect the cubic factor in \eqref{fdef} which is needed to derive \eqref{tbound}.}
\begin{equation}\label{fdef}
f_{\lambda_1 \lambda_2}(t) = \frac{(\Lambda^2-2m^2+t/2)^3}{(\Lambda^2-2m^2)^3} \partial_s^2 {\cal A}_{\lambda_1 \lambda_2}(s,t) |\Big|_{s=2m^2-t/2} \, , 
\end{equation}
then assuming that the branch cut in $s$ begins at $\Lambda^2$\footnote{This can always be arranged by using the improved positivity bounds to remove the low-energy part of the cut.} the following bound can be derived from the dispersion relation:
\begin{equation} \label{tbound}
\frac{|f_{\lambda_1 \lambda_2}(t)|}{f_{\lambda_1 \lambda_2}(0)} \le 1+ \mathcal{O}\!\left(\frac{\sqrt{|t|m}}{\Lambda^2}\right) \, , 
\end{equation}
where $\Lambda$ is the assumed cutoff of the EFT. 
Ref.~\cite{Bellazzini:2023nqj} then applies this to the scattering amplitudes of dRGT for $t<0$ and $|t|$ comparable to $\Lambda^2$ which is well outside the physical region and the largest magnitude at which we can feasibly trust the EFT. From this \cite{Bellazzini:2023nqj} infers the strong conclusion that massive gravity only remains consistent with positivity bounds, and hence a local UV completion, if 
$\Lambda \sim {\cal O}(30) m$, which is a sufficient hierarchy to consider the EFT meaningful since $m^2/\Lambda^2 \sim 10^{-3}$, but is clearly a problem for phenomenological applications. 

 If correct, one may argue that this result is rather unsurprising on these scales. Indeed, with a large hierarchy $|t| \gg m^2$, we may reasonably trust the Galileon decoupling limit of massive gravity, with corrections to that limit being sufficiently suppressed by the graviton mass $m$. The Galileon, with a small mass term, is known to be inconsistent with positivity bounds \cite{Tolley:2020gtv}. However, as discussed below, the conclusions of \cite{Bellazzini:2023nqj} are in fact only valid under overly-restrictive tree-level assumptions. Without relying on this implicit (and typically unexpected) assumption, the presented analysis remains inconclusive at best in its current form. 
 The main issue with the analysis of \cite{Bellazzini:2023nqj} is that the derivation of \eqref{tbound} relies on the partial-wave expansion and on unitarity, yet the resulting relation is subsequently applied outside the region where the unitarity relation is valid. While one may still assume the existence of a dispersion relation, its discontinuity can no longer be inferred from unitarity. Consequently, \eqref{tbound} can only be rigorously justified for $-4m^2 < t \le 0$, since only within this range can the relevant discontinuity be determined. For instance, even if loop corrections remain perturbatively small in the physical region, they generate a branch cut beginning at $|t|=4m^2$. The associated non-analyticity then precludes any conclusion regarding the size of loop corrections at values of $|t|$ comparable to $\Lambda^2$.

Put differently, the issue is not that the amplitude itself cannot be reliably computed in the regime $m^2<-t<M^2$.
Rather, the problem is that the dispersion relation cannot be extended into this domain. Analyticity is only established within the region $-4m^2 < t < 4m^2$, and not necessarily beyond it. The argument therefore implicitly relies on a perturbative expression whose derivation assumes analyticity in a region where analyticity has not been demonstrated and is, in fact, expected to fail. From this perspective, the conclusions of \cite{Bellazzini:2023nqj} are avoided simply because analyticity need not be assumed in that region to begin with.

One may nevertheless contemplate the possibility that massive gravity admits a purely tree-level UV completion. A familiar example is provided by perturbative string theory, where the Veneziano and Virasoro-Shapiro amplitudes, together with their generalizations, provide explicit tree-level UV completions of Yang-Mills and supergravity scattering amplitudes. In such cases, dispersive arguments can consistently be applied directly to the tree-level amplitudes.

Under the assumption of a tree-level UV completion, loop effects are absent and therefore no branch cuts are generated. It is then plausible that the dispersion relation remains valid throughout the spacelike region $t<0$, up to values $|t|\sim \Lambda^2$, where $\Lambda$ now denotes the mass of the lightest state, other than the graviton, appearing in the UV completion. In that case, \eqref{tbound} may legitimately be interpreted as a constraint on the tree-level amplitude of the putative UV completion. The appropriate conclusion of \cite{Bellazzini:2023nqj} is therefore more modest: if massive gravity admits a tree-level UV completion, then the mass of the lightest state beyond the EFT is constrained to be of order ${\cal O}(30)\,m$. However another possibility is that massive gravity does not admit a tree-level UV completion.

 It should be stressed that there is no expectation that massive gravity admits a tree-level UV completion, rather the opposite. From inception, it was expected that any theory that exhibits the Vainshtein mechanism in a consistent way must necessarily be strongly coupled on the scale $\Lambda_3$ or something comparable. This is an integral part of the redressing process that must take place when considering the Vainshtein mechanism in the background of matter sources \cite{deRham:2014wfa}. 
 
 Loop corrections may be small and perturbative in the physical region, but we cannot a priori assume that they are small as the scattering amplitude is extended away from this. In fact, the arguments of \cite{Bellazzini:2023nqj} rely on the observation that if one were to push the inequality to larger $|t|$, it would lead to ever more stringent bounds. This means that the bound is dominated by the region where the amplitude is least under control. What can be conclusively argued from this observation is that massive gravity does not admit a UV completion where it would be safe to assume the dispersion relation with the discontinuity fixed by unitarity for $|t|$ all the way to the cutoff, unless that cutoff is parametrically close to the graviton mass. Thus, it is quite reasonable to expect that the dispersion relation, whose discontinuity is constrained by unitarity, ceases to provide a good approximation beyond a scale of $|t|$ that is parametrically tied to the graviton mass, precisely as one would naturally anticipate in any quantum theory. While this has been thoroughly scrutinised, it is interesting to note that to date arguments presented in the literature do not {\it yet} preclude the existence of a consistent {\bf unitary} and {\bf local} UV completion, although future work may sharpen this conclusion. 
 
\paragraph{Massive Gravity @ 15:} It remains an open and highly active question whether massive gravity is consistent with positivity bounds and there is as yet no proof that it is inconsistent with a local UV completion. It is of course possible that massive gravity ultimately proves to be inconsistent with positivity bounds and hence does not admit a local, analytic, Lorentz-invariant UV completion. Most likely this is indicative of a {\bf non-local} UV completion that is otherwise Lorentz invariant and analytic, and we will see evidence for this below.

Putting these considerations aside, it is interesting to explore whether there is a controlled situation in which the UV properties of massive gravity can be understood. Fortunately, there is at least one situation where this is true, massive gravity in two dimensions, which we shall consider in the next section.

\subsection{Massive Gravity as a $T\bar T$ Deformation}
\label{sec:TTbar}

In recent years, it has been recognised that massive gravity in two dimensions can in fact be exactly solved, for suitably chosen matter, and admits a well-defined non-perturbative description. This follows from the observation, first made in \cite{Tolley:2019nmm}, that two-dimensional massive gravity is, at both the classical and quantum mechanically equivalent to a particular modification known as a $T\bar T$ deformation (see also \cite{Mazenc:2019cfg,Tsolakidis:2024wut,Nix:2025plr}). Before explaining this connection, we will first review these deformations. 

\subsubsection*{$T\bar T$ Deformations}

It is generically the case that adding an irrelevant operator to an otherwise renormalizable or finite quantum field theory will render it non-renormalizable, demanding it be regarded as an EFT and requiring the introduction of an infinite number of other irrelevant operators to ensure renormalizability at all orders. There is, however, in two spacetime dimensions a fascinating exception to this: the so-called $T\bar{T}$ deformation, which defines a continuous family of theories obtained by deforming a seed field theory in a highly constrained manner \cite{Zamolodchikov:2004ce}. In most cases, the seed field theory is regarded as conformal or integrable, although the equivalence extends beyond this.

The $T\bar{T}$ deformation is defined through the flow equation
\begin{equation} \label{TTbar}
\frac{\partial S_\lambda}{\partial \lambda}
=
- \int \d^2x\, \det T_{\lambda},
\end{equation}
where $S_\lambda$ denotes the deformed action for coupling $\lambda$, and the composite operator $\det T_{\lambda}$ is a function of the stress-energy tensor at that particular value of $\lambda$ (\ie the stress tensor of $S_{\lambda}$) which in two dimensions is
\begin{equation}
\det T
=
-\frac{1}{2}
\left(
T_{\mu\nu}T^{\mu\nu} - T^2
\right).
\end{equation}
The name is a reflection of the fact that for a conformal field theory for which $T^{\mu}_{\mu}=0$, this evaluates to $T_{++} T_{--}=T \bar T$ in lightcone coordinates.
Remarkably, if the seed theory is integrable, then this deformation preserves the integrability properties of the original theory and allows an exact determination of the finite-volume spectrum and scattering data, despite introducing an infinite tower of higher-dimensional operators \cite{Smirnov:2016lqw,Cavaglia:2016oda}. 

Since \eqref{TTbar} is a flow equation, integrating to finite values of $\lambda$ will generate an infinite number of terms.
The most striking feature is that despite this, the effect of the deformation on the full $S$-matrix is given by a simple, universal phase factor,
\begin{equation}
S(\{p_i\})
\;\longrightarrow\;
\left[
\prod_{i<j}
e^{\frac{i}{2}\lambda\,\epsilon_{ab}\,p_i^a p_j^b}
\right]
S(\{p_i\}) ,
\end{equation}
or for $2\to2$ scattering,
\begin{equation}
e^{2i\delta(s)}
=
e^{\frac{i}{2}\lambda s}
e^{2i\delta_0(s)} .
\end{equation}
This shows that the deformation does not modify the magnitude of the amplitude but dresses it with an energy-dependent phase that grows linearly with the Mandelstam variable $s$ \cite{Dubovsky:2017cnj}.

We may regard $T\bar{T}$ as perhaps the simplest example of a new class of non-Wilsonian field theories. They are non-Wilsonian in the sense that despite being naively non-renormalizable, indeed containing an effective action with an infinite number of local terms, they are nevertheless meaningful and are not defined by a fixed point at high energies. Inspection of their features indicates that they are more non-local than ordinary field theories and thus belong to the Jaffe class of nonlocalizable field theories. This non-locality is indicated by the exponential growth of the scattering amplitude in certain directions, violating the polynomial boundedness or exponential boundedness $|{\cal A}(s)|<|s|^N e^{\alpha \sqrt{|s|}}$.
This behaviour reminds us of what we might expect for a gravitational theory, where the absence of local gauge-invariant observables suggests the breakdown of strict locality at high energies, \ie a violation of polynomial or exponential boundedness.

\subsubsection*{Equivalence with massive gravity}

By itself, the $T \bar T$ deformation is a deformation of a non-gravitational theory and appears to have nothing to do with gravity at all. To see the first hints of the connection, it is helpful to consider the tree-level exchange amplitude of a massive spin--2 field between two conserved sources $T_{\mu\nu}$ and $\tilde{T}_{\mu\nu}$ in $D$ spacetime dimensions:
\begin{equation}
\mathcal{A}_{TT}
=
\frac{1}{M_{\mathrm{Pl}}^{D-2}}
\int \d^D x \d^D y
\left(
T_{\mu\nu}(x)\tilde{T}^{\mu\nu}(y)
-
\frac{1}{D-1}T(x)\tilde{T}(y)
\right)
(\Box - M^2 + i\epsilon)^{-1}(x,y),
\end{equation}
where $M$ denotes the mass of the exchanged spin--2 field. The specific tensor structure reflects the projection onto the physical polarisations of a massive graviton, and the relative coefficient between the traceless and trace parts is fixed by Lorentz invariance and consistency with a massive spin--2 representation.
In the limit $D \to 2$ we have
\cite{Tolley:2019nmm}
\begin{equation}
\det T
=
-\frac{1}{2}
\lim_{D \to 2}
\left(
T_{\mu\nu}T^{\mu\nu}
-
\frac{1}{D-1}T^2
\right),
\end{equation}
which precisely reproduces the composite operator appearing in the definition of the $T\bar{T}$ deformation \cite{Zamolodchikov:2004ce}. 
By itself, this argument is insufficient, as the deformation is actually defined by a flow equation, and it is not clear that this perturbative correction will match the deformation exactly. However, in \cite{Tolley:2019nmm} it is shown that by formulating two-dimensional massive gravity in the vielbein formalism and minimally coupling it to the field theory (\eg a CFT), the act of integrating out the vielbein/metric gives an exact solution to the flow equation that defines the deformation, and this holds both at the classical and quantum level. Consequently, ghost-free (dRGT) massive gravity in two spacetime dimensions coupled to a field theory is exactly equivalent to a $T \bar T$ deformation of the field theory\footnote{It was previously noted that there is a connection to a version of JT gravity \cite{Dubovsky:2017cnj}. However, the equivalence here with massive gravity is more direct and hence more illuminating.} \cite{Tolley:2019nmm,Mazenc:2019cfg}. This explains at least in part why the $T \bar T$ deformation of a non-gravitational field theory has properties in common with a gravitational theory. 

The solvability of the deformation in $D=2$ reflects the absence of local propagating graviton degrees of freedom, allowing the gravitational interaction to be resummed exactly into a universal phase dressing of the scattering matrix \cite{Smirnov:2016lqw,Dubovsky:2018bmo}. Indeed, we may be forgiven for thinking that there is no such thing as massive gravity in two dimensions, after all the Einstein-Hilbert term is topological and carries no local dynamics. While it is true that there are no propagating degrees of freedom, there are still the static aspects of the gravitational forces, and the scattering of matter is affected by the gravitational dynamics. 

There are interesting relations between massive gravity and holography \cite{McGough:2016lol,Hartman:2018tkw,Guica:2019nzm,Gorbenko:2018oov}, and attempts have been made to generalise this connection to higher dimensions \cite{Taylor:2018xcy}. Recent work has made this connection in more general dimensions more precise \cite{Tsolakidis:2024wut,Nix:2025plr}.
These considerations were further explored in \cite{Nix:2025plr} where the connection between the trace-flow equation and the local renormalisation group is clarified in arbitrary dimensions. Remarkably,  the usual initial condition for the coupling leads to the minimal dRGT potential in any dimension. 

 These findings support viewing massive gravity and related infrared deformations of gravity as non-local yet self-consistent theories, just with very different ultraviolet properties due to the gravitational dressing. They further imply that even if massive gravity is shown to violate the positivity bounds expected for a local UV completion, these violations are benign, merely reflecting that massive gravity is intrinsically gravitational and therefore more fundamentally non-local \cite{Dvali:2012zc,Dubovsky:2012wk,Dubovsky:2013ira,Dubovsky:2017cnj}. A similar statement may be expected to hold for Galileon theories \cite{Keltner:2015xda} and see \cite{Aydemir:2012nz} for how resonant structures and strongly coupled effects can lead to a self-healing of unitarity.

\subsection{Islands and Massive Gravity}

Recent work on the unitary evolution of black holes and the derivation of the Page curve within semiclassical calculations has suggested the need to include `islands', spacetime regions necessary to correctly compute the entanglement entropy of black hole radiation \cite{Penington:2019npb,Almheiri:2019psf,Almheiri:2019hni}.
Ref.~\cite{Geng:2021hlu} argues that there is a structural obstruction to interpreting islands as consistent entanglement wedges in theories with strictly massless gravity. The long-range gravitational Gauss laws \cite{Donnelly:2016rvo} required by a theory of massless gravity prevent operator factorisation across the putative island region and its complement. In known controlled higher-dimensional constructions where islands are realised, this obstruction is avoided due to the emergence of an effective graviton mass \cite{Geng:2020qvw,Geng:2023zhq,Geng:2025byh}. The most well-known example in which gravity emerges as effectively massive in AdS is the Karch-Randall model \cite{Karch:2000ct,Karch:2001cw,Porrati:2001gx}.

In \cite{Cao:2021ujs}, island prescriptions are applied directly to black hole solutions in dRGT massive gravity. The resulting entanglement entropy exhibits a Page-curve behaviour consistent with expectations\footnote{Interestingly, the reference metric considered in \cite{Cao:2021ujs} is two-dimensional, as was considered also in other examples related to holography \cite{Alberte:2013sma}.}. This provides a concrete illustration of how a graviton mass can regulate information recovery in semiclassical black hole settings, and also demonstrates how in one aspect massive gravity is an improvement on GR. In \cite{Demulder:2022aij}, islands are studied in a type IIB string theory realisation of AdS$_4$ massive gravity coupled to a bath. 
In \cite{Miao:2022kve}, island constructions in wedge holography with DGP-type modifications are revisited, and it is found that reproducing island behaviour in the massless limit can be accompanied by instabilities, such as ghost-like modes. 

It is fair to say that the relevance of massive gravity theories to entanglement entropy calculations is still very much an open question, and it may only be that these calculations are better defined in massive gravity for the same reason that scattering amplitudes are better defined in massive gravity as they do not exhibit infrared divergences. Nevertheless, this is a remarkable connection; see \cite{Geng:2025rov} for recent progress.

\vspace{0.3cm}

\noindent\textbf{\color{color1}Bottom Line:}
In some cases, Ghost-free massive gravity admits a manifestly dynamical formulation in which the full constraint structure can be exhibited explicitly. In parallel, its connection with  $T\bar{T}$-type deformations may provide a partially solvable window some of its UV behavior. This may suggest that massive gravity can be in principle be regarded as a theoretically controlled IR modification of gravity, although it may be expected that its effective description becomes non-local near its strong coupling scale.

\section{Conclusions and Outlook}

Lorentz-invariant, ghost-free massive gravity (dRGT) constitutes a rare and instructive case of an infrared modification of gravity that remains consistent and predictive within its domain of applicability. Its construction 15 years ago resolved long-standing conceptual obstacles associated with interacting massive spin--2 fields. It shows that General Relativity allows nontrivial, well-controlled deformations that, with only a few free parameters, are essentially as unique as GR itself. The key achievement was the identification of a highly constrained nonlinear interaction structure that propagates exactly five degrees of freedom, avoids the Boulware-Deser ghost beyond the linear regime, and raises the cutoff of the EFT to the highest possible scale.

Because massive gravity contains more propagating degrees of freedom than GR, it admits a wealth of solutions that go beyond simple deformations of those in Einstein gravity. This, in turn, gives rise to a rich landscape of phenomenological and cosmological possibilities that remain only partially explored. Although strictly spatially flat FLRW solutions are excluded in the most straightforward implementations, this should be seen as an advantage, as it is an essential ingredient of the degravitation framework. Consistent alternatives are available, such as inhomogeneous cosmological models—where the graviton’s Compton wavelength sets the coherence scale—as well as open-Universe constructions. The rigid framework of Lorentz-invariant massive gravity can easily be generalised, for example, to bigravity \cite{Hassan:2011zd}, generalised massive gravity, Lorentz violating massive gravity \cite{deRham:2014zqa,deRham:2023byw} and also motivated nontrivial extensions to interacting spin-1 theories \cite{deRham:2020yet}.

Massive gravity remains viable because agreement with local tests of gravity is ensured through the Vainshtein mechanism \cite{Vainshtein:1972sx,Babichev:2013usa}. The Vainshtein mechanism dynamically suppresses deviations from General Relativity near regions of large energy density, \ie matter sources. The robustness of this mechanism has been confirmed through analytic arguments and fully nonlinear numerical simulations \cite{Gerhardinger:2024rza,deRham:2024xxb}.

In the last few years, substantial progress has also been achieved in formulating massive gravity as a well-posed dynamical system \cite{Kozuszek:2024vyb,Kozuszek:2025lem,Albertini:2024kmf}. Very recently, new analytic results have shown that helicity-2 gravitational waves always travel along the light cone (at high frequencies), even on arbitrary backgrounds, offering a striking fully non-linear confirmation with important consequences for upcoming gravitational-wave surveys \cite{deRham:2026zqu}.

The vielbein-based formalism appears to be particularly useful in this regard since it leads to an algebraic form for the scalar constraint, which facilitates transparent degrees-of-freedom counting and is readily implementable numerically.
These developments open the door to fully nonlinear simulations of strong-field and time-dependent phenomena, including gravitational-wave dynamics, and provide answers for the more nontrivial dynamical questions that have so far proven elusive due to the complicated nature of the equations of motion. Connections between these developments and phenomenology, including binary mergers, binary pulsars, and primordial gravitational waves would be particularly exciting to explore. 

Ghost-free massive gravity is intrinsically an EFT with close connections to the class of Galileon theories. Accordingly, it remains an open question whether massive gravity admits a UV completion. There has been considerable work in the last 15 years using the ideas of positivity bounds/S-matrix bootstrap methods to put constraints on massive gravity. Interestingly, these constraints appear to favour ghost-free models with their $\Lambda_3$ coupling scale. The question of whether there exists a local UV completion has not yet been fully resolved, although remarkable progress has been made.

In addition to being a phenomenologically interesting foil to GR, massive gravity shows up in a number of different areas which touch on our fundamental understanding of quantum gravity. More precisely, two-dimensional massive gravity minimally coupled to a field theory, far from trivial, is exactly equivalent to a $T \bar T$ deformation of that field theory. This gives a new perspective on its possible UV completion and high-energy properties. Massive gravity has also emerged in discussions on islands in calculations of entanglement entropy and the Page curve for black holes using holographic ideas.

\vspace{1cm}
\noindent {\bf Acknowledgments:}
CdR would like to thank  Andrew Tolley for useful discussions, comments, and suggestions. The work of CdR is supported by STFC Consolidated Grant ST/X000575/1 and by a Simons Investigator award 690508. CdR is also grateful for support from the Foundational Questions Institute (FQxI).

\bibliographystyle{JHEP}
\bibliography{references.bib}

\end{document}